\documentclass[twocolumn]{aastex631}
\usepackage{comment}
\usepackage{natbib} 
\usepackage{amsmath}
\usepackage[mathlines]{lineno}

\newcommand{\kms}{km s$^{-1}$}

\newcommand{\hi}{H~{\scriptsize I}}
\newcommand{\hii}{H~{\scriptsize II}}
\newcommand{\cii}{C~{\scriptsize II}}

\newcommand{\degree}{$^{\circ}$}

\usepackage{graphicx} 
\usepackage{multirow}

\begin{document}

\defcitealias{Butterfield2024}{FIREPLACE I}
\defcitealias{Butterfield2024ring}{FIREPLACE II}
\defcitealias{Pare2024}{FIREPLACE III}
\defcitealias{Pare2024b}{FIREPLACE IV}
\newcommand{\rz}[1]{{\bf \color{cyan}{[note: #1]}}}
\newcommand{\mm}[1]{{\bf \color{blue}{[MM: #1]}}}

\title{SOFIA/HAWC+ Far-Infrared Polarimetric Large Area CMZ Exploration Survey. V. The Magnetic Field Strength and Morphology in the Sagittarius C Complex}

\correspondingauthor{Roy J. Zhao}
\email{rzhaolx@uchicago.edu}

\author[0009-0001-9716-4188]{Roy J. Zhao}
\affiliation{Kavli Institute for Cosmological Physics, The University of Chicago, 5640 S Ellis Ave., Chicago, IL 60637, USA}
\affiliation{Department of Physics, The University of Chicago, 5720 S Ellis Ave., Chicago, IL 60637, USA}
\affiliation{Department of Physics \& Astronomy, University of California, Los Angeles, 475 Portola Pl., Los Angeles, CA 90095, USA}

\author[0000-0002-6753-2066]{Mark R. Morris}
\affiliation{Department of Physics \& Astronomy, University of California, Los Angeles, 475 Portola Pl., Los Angeles, CA 90095, USA}

\author[0000-0003-0016-0533]{David T. Chuss}
\affiliation{Department of Physics, Villanova University, 800 E. Lancaster Ave., Villanova, PA 19085, USA}

\author[0000-0002-5811-0136]{Dylan M. Par\'e}
\affiliation{Department of Physics, Villanova University, 800 E. Lancaster Ave., Villanova, PA 19085, USA}

\author[0000-0001-8819-9648]{Jordan A. Guerra}
\affil{Cooperative Institute for Research in Environmental Sciences, University of Colorado, Boulder, CO 80309 USA}

\author[0000-0002-4013-6469]{Natalie O. Butterfield}
\affiliation{National Radio Astronomy Observatory, 520 Edgemont Rd., Charlottesville, VA 22903, USA}

\author[0000-0002-7567-4451]{Edward J. Wollack}
\affiliation{NASA Goddard Space Flight Center, Mail Code: 665, Greenbelt, MD 20771, USA}

\author[0009-0006-4830-163X]{Kaitlyn Karpovich}
\affiliation{Kavli Institute for Particle Astrophysics \& Cosmology, P.O. Box 2450, Stanford University, Stanford, CA 94305, USA}
\affiliation{Department of Physics, Stanford University, Stanford, CA 94305, USA}
\affiliation{Department of Physics, Villanova University, 800 E. Lancaster Ave., Villanova, PA 19085, USA}


\begin{abstract}
    We present an analysis of the magnetic field strength and morphology in the Sagittarius C complex (Sgr C; G359.43-0.09) in the Milky Way Galaxy's Central Molecular Zone (CMZ), using the 214 \micron{} polarimetry data acquired with the High-resolution Airborne Wide-band Camera (HAWC+) instrument aboard the Stratospheric Observatory for Infrared Astronomy (SOFIA). We conduct a modified Davis-Chandrasekhar-Fermi (DCF) analysis of individual clouds and find that the sky-plane magnetic field strength varies from highly turbulent regions having inferred strengths of $\sim30~\mu{\rm G}$ to regions of relatively uniform field orientation having strengths of $\sim 300~\mu{\rm G}$. Several hundred magnetic field pseudovectors in the Sgr C region were measured to trace the projected magnetic field orientation within cold molecular clouds, and as is the trend throughout the CMZ, they show a higher polarization fraction toward the periphery of the clouds. The magnetic field orientations suggest that outflows from active star-forming regions, such as the G359.43-0.10 extended green object (EGO) and the protostellar source FIR-4 (G359.43+0.02), cause high turbulence in their vicinity. The magnetic field direction is found to be tangential to the surface of the Sgr C \hii{} region, which displays spatial correspondence with two [\cii{}] emission cavities reported in the \hii{} region, signifying a compression front between the \hii{} region and the surrounding dense clouds. Several other features in the vicinity of Sgr C, especially numerous non-thermal radio filaments (NTFs) and a diffuse source of X-ray emission to the immediate southwest of the \hii{} region, are discussed with regard to the magnetic field measurements.
\end{abstract}

\section{Introduction}
The central half kiloparsec of the Milky Way disk, commonly referred to as the Central Molecular Zone (CMZ), is an ideal laboratory to study astrophysics in extreme environments given its high density, massive star-formation, and complex structure \citep[see][and references within]{Morris1996, Ferriere2007, Bryant2021, Henshaw2023}. Apart from the extensively studied star formation and Interstellar Medium (ISM) properties, a strong magnetic field has also been found to permeate the CMZ \citep{Tsuboi1985, LaRosa2005, Morris2006, Ferriere2009, Morris2015, Akshaya2024, Pare2024, Tress2024}, especially in dense clouds \citep{Yusef-Zadeh1999, Pillai2015, Hsieh2018, Butterfield2024, Lu2024}. The relatively strong magnetic field (0.1 -- 1 mG) is said to counteract the gravitational collapse of clouds and contributes to the lower than expected star-formation rate in this region given its high density \citep{Morris1993, Chuss2003, Kruijssen2014, Morris2023}, therefore of great interest to be studied. 

In this paper, we use the results of the Far-Infrared Polarimetric Large Area CMZ Exploration survey (FIREPLACE; PI: D. Chuss), a high-resolution dust polarimetry survey conducted at 214 \micron{}, to study the magnetic field in the CMZ. As a common experimental probe of the sky-plane component of the magnetic field, dust polarimetry exploits the property that the spin axes of dust grains tend to be aligned with the magnetic field \citep{Andersson2015}. The measurement of polarized emission from clouds therefore reveals the alignment of dust grains within, indicating the sky-plane magnetic field component integrated along the line of sight \citep{Chuss2003, Andersson2015}. 

There are several possible mechanisms that could be responsible for the field orientations deduced from dust polarimetry observations. As the magnetic field lines are anchored inside the molecular clouds, the field lines could be sheared by the clouds' internal motions \citep{Parker1979}. In a differentially orbiting medium like the CMZ, the shear stretches the magnetic field into a primarily toroidal geometry observed in large scale \citep{Werner1988, WardleKonigl90}. The same effect can also be indicative of the direction of the internal shear within clouds on a smaller scale \citep{Aitken1991, Morris1992, Aitken1998, Novak2000, Chuss2003, Morris2015}. Other candidates that can contribute to the magnetic field orientations, such as cloud compression, which occurs at the interface of expanding \hii{} regions or supernova blast waves, tend to shape the field to be parallel to the compressive interface. 

Indeed, a high-resolution magnetic field study, like the FIREPLACE survey, can benefit our understanding of many aspects of the CMZ, such as the magnetic field morphology within molecular clouds and within portions of low-density ISM occupied by the NTFs \citep{Yusef-Zadeh1984, Morris1985, Heywood2022, Yusef-Zadeh2022index}. These NTFs are manifested as spatially extended synchrotron emission from relativistic electrons constrained to diffuse along a magnetic flux tube. The NTFs within the CMZ are generally oriented perpendicular to the Galactic plane, which implies a large-scale vertical magnetic field pervading the Galactic Centre (GC), thereby pointing to the existence of an organized inter-cloud magnetic field \citep{Morris1996, Heywood2022}. Furthermore, \citet{Serabyn1994} argued that turbulent instabilities at the ${\rm H_2}$/\hii{} interface could mix two orthogonal magnetic field systems inside and outside that interface, which facilitates magnetic field line reconnection. Such phenomenon can boost free electrons in the ISM to relativistic velocities, thereby forming NTFs as those electrons diffuse along the magnetic field lines in the inter-cloud ISM \citep{Serabyn1994, Morris1996IAUS, Chuss2003}. In addition, many other hypotheses have been offered to account for the NTFs \citep[see references in][]{Morris1996IAUS, Boldyrev2006, Yusef-Zadeh2004, BB+16, BB+18, YZW19}. For example, an alternative hypothesis for generating relativistic electrons that does not directly involve magnetic fields invokes diffusive shock acceleration where stellar winds from massive young stars impact relatively dense interstellar material \citep{Rosner1996}. 

Previous studies of the GC magnetic field, including those using the Submillimeter Polarimeter for Antarctic Remote Observations (SPARO; \citealt{Novak2003}), {\it Planck} \citep{Planck2015}, the Polarized Instrument for the Long-wavelength Observation of the Tenuous ISM (PILOT) \citep{Mangilli2019}, and the Atacama Cosmology Telescope (ACT; \citealt{Guan2021}) have focused on the large-scale magnetic field, with typical resolutions $\gtrsim 1\arcmin{}$. These works reported a largely uniform magnetic field that permeates the CMZ, and is oriented at a $\sim 20$\degree{} angle with respect to the Galactic plane. However, conclusions made from these surveys remain on a global level, since the spatial resolution of these surveys does not generally resolve the internal structure of individual molecular clouds. High-resolution surveys conducted on subregions within the CMZ, such as those reported by \citet{Chuss2003} and \citet{Lu2024} found a correlation between the magnetic field orientation and molecular cloud density, though they do not have enough coverage to cover a sufficiently large portion of the CMZ to draw global conclusions. Along the same line in Sgr C, \citet{Bally2024} studied the magnetic field inside the prominent \hii{} region using the JWST-NIRCam images. With the FIREPLACE survey dataset, a recent study by \citet{Pare2024b} (or \citetalias{Pare2024b}) demonstrated the advantage of studying the magnetic field orientation in the CMZ using datasets with both high resolution and large coverage. 

\begin{figure}
    \centering
    \includegraphics[width=\columnwidth]{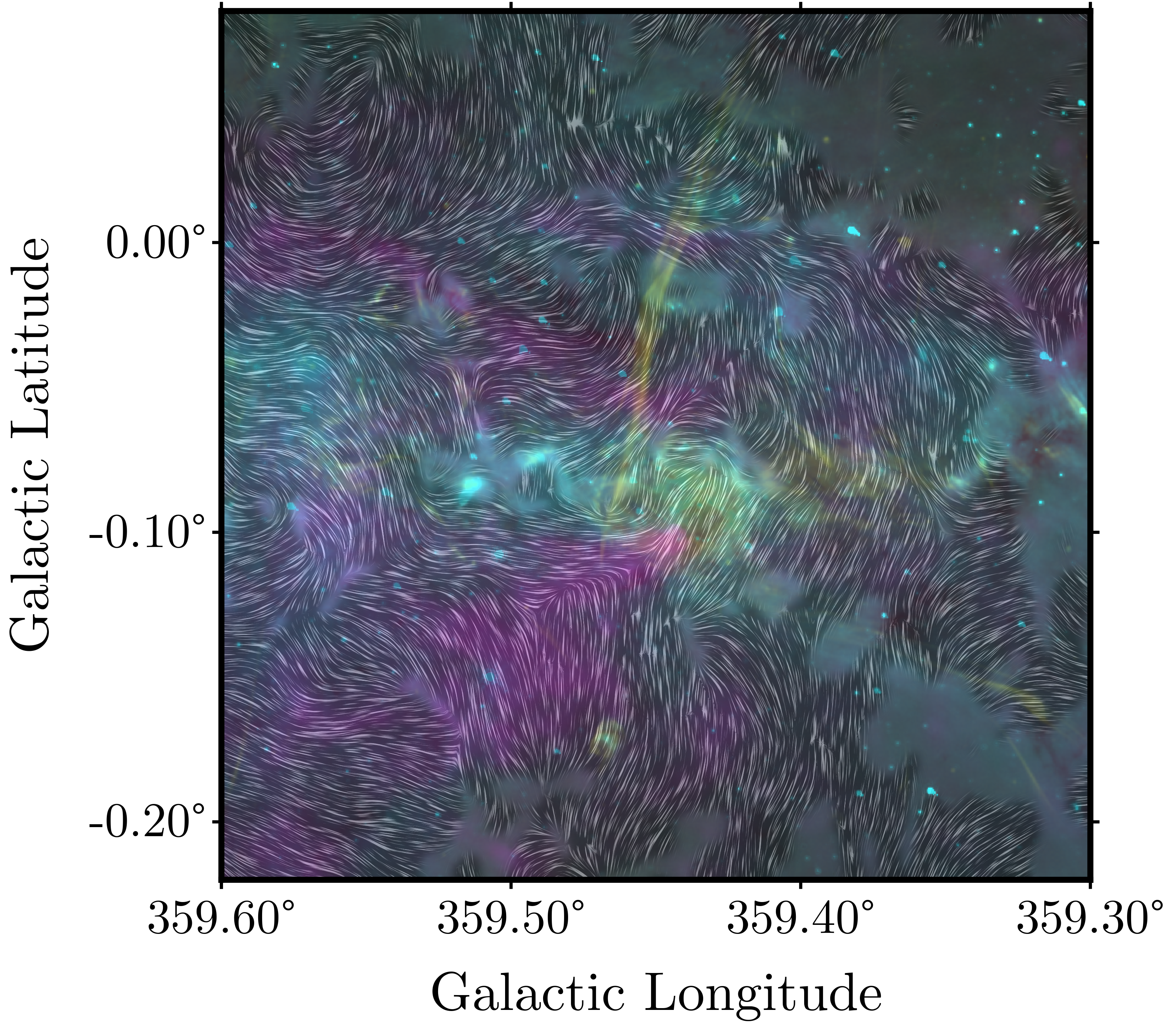}
    \caption{A tri-colour view of the Sgr C complex with the {\it Spitzer}/IRAC 8 \micron{} image in cyan \citep{Stolovy2006}, the {\it Herschel}/SPIRE 250 \micron{} image in magenta \citep{Molinari2010}, and the MeerKAT/L-Band 20 cm image in yellow \citep{Heywood2022}, overlaid with a LIC rendition of the magnetic field measurements reported by \citet{Pare2024}. The main structures, including the Sgr C \hii{} region, the Sgr C NTF, FIR-4, and the 100 pc ring, are highlighted in Figure~\ref{fig:SgrC_finding}. }
    \label{fig:SgrC}
\end{figure}

\begin{figure}
    \centering
    \includegraphics[width=\columnwidth]{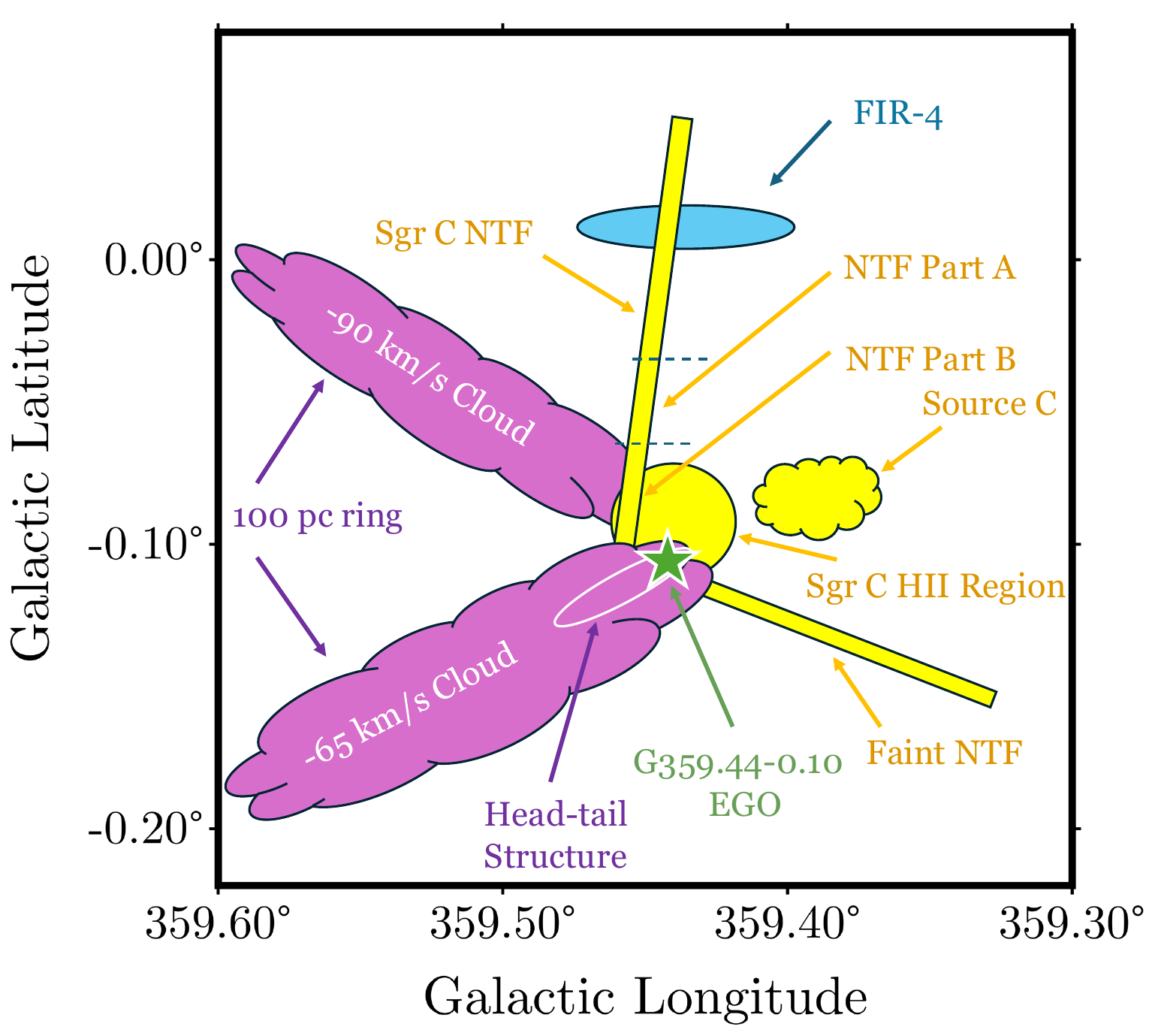}
    \caption{Schematic diagram of the main features within the Sgr C complex on the sky-plane, including FIR-4 (blue ellipse), the major molecular clouds detected in far-IR (purple clouds denoted by their measured velocities), the Sgr C \hii{} region and NTF detected in radio continuum (yellow circle and long, vertical rectangle, respectively), the G359.44-0.10 EGO (\citealt{Kendrew2013}; green star), and the radio and mid-IR structure known as source C (yellow cloud). The velocities of the molecular clouds depicted in magenta are obtained using the molecular line observation of the 3 mm Mopra survey reported by \citet{Jones2012}, detailed in Section~\ref{sec:spectral}.}
    \label{fig:SgrC_finding}
\end{figure}

In this paper, we focus on studying the magnetic field within one of the sub-regions in the CMZ --- the Sgr C complex. Sgr C was initially reported as one of the most prominent \hii{} regions in the GC \citep{Downes1966}, and it was later detailed by \citet{Liszt1985b} and \citet{Liszt1995}, who noted the presence of a prominent NTF apparently connected to the \hii{} region. Figure~\ref{fig:SgrC} portrays this region at multiple wavelengths overlaid with a line integral contour (LIC; \citealt{Cabral1993}) image to represent the magnetic field direction made by measurements reported by \citet{Pare2024} (or \citetalias{Pare2024}).
Figure~\ref{fig:SgrC_finding} shows a simplified schematic diagram of Sgr C to highlight the important structures within.

The large-scale astrogeography of Sgr C is also of particular interest and, may be crucial in reaching a full understanding of the global structure of the CMZ. Many studies found that the Sgr C complex resides on the projected western vertex of a dust ring associated with the $X_2$ orbit due to the Galactic bar potential \citep{Molinari2011, Krumholz2015, Sormani2018, Salas2020}. Through the lens of simulations, \citet{Sormani2024} and \citet{Tress2024} alternatively described the formation of the ring as a result of the particular gas dynamics in the Galactic bar, where shocks and magnetic field are also considered in these computations. Therefore, this theory places Sgr C and its positive-longitude counterpart, Sgr B2, at the two interaction sites of $X_1$-$X_2$ orbits, located at both projected ends of the nuclear stellar disk. Meanwhile, the Mopra survey \citep{Jones2012} revealed a significant amount of shock tracers (SiO and HNCO) within the Sgr C complex. The widely distributed shock tracers are presumably induced by the $X_1$-$X_2$ orbit collision. The exact role of this collision is yet unclear since it can either compress the clouds and provoke star formation, or increase turbulence and thereby inhibit star formation. As magnetic field measurements could reveal information about the large-scale turbulence of dense clouds, a thorough study of the magnetic field morphology within the Sgr C complex may provide improved characterization of a region subjected to large-scale shocks and how star formation enters the picture in such a region. 

In this paper, we study the magnetic field measurements from the FIREPLACE survey within the Sgr C complex, in the context of other known structures in this region. The paper is organized as follows. In Section~\ref{sec:observations}, we briefly describe the FIREPLACE survey. In Section~\ref{sec:bfield}, we present analyses of the Sgr C magnetic field measured in 214 \micron{}. In Section~\ref{sec:discussion}, we use multi-wavelength observations to investigate numerous intriguing structures in the Sgr C region where the magnetic field has apparently become manifest. In Section~\ref{sec:features}, we present a detailed examination of several features in which the magnetic field is likely playing an important role. We finally summarize our conclusions in Section~\ref{sec:conclusion}. Throughout this paper, we adopt 8.2 kpc as the distance to the GC \citep{GRAVITY2019}.

\section{Observations and Data}
\label{sec:observations}
The 214 \micron{} intensity and polarization results presented in this paper utilize the second data release (DR2) of the FIREPLACE survey described in \citet{Pare2024}. The data were acquired using HAWC+ \citep{Harper2018}, an imaging polarimeter aboard the SOFIA telescope. The complete survey, including the first data release (DR1; \citealt{Butterfield2024}, or \citetalias{Butterfield2024}) covers the inner 1.5 degrees of the CMZ with a spatial resolution of 19.6\arcsec{}. We refer interested readers to \citet{Pare2024} for full details of the observations and data reduction.

In this paper, the inferred magnetic field orientations represent the component of the magnetic field projected onto the plane of the sky and integrated along the line of sight. Additionally, we assume that the B-RAT dust grain alignment mechanism dominates in the Sgr C complex. Therefore, the magnetic field orientations are obtained by rotating the polarization orientations by 90 degrees \citep{Andersson2015}. For a full visual representation, the length of the magnetic field pseudovectors presented in this paper is scaled to be proportional to the debiased polarization fraction, defined by 
\begin{equation}
    p=\sqrt{p_m^2-\sigma_p^2},
\end{equation}
where $p_m$ is the measured polarization fraction and $\sigma_p$ is the associated uncertainty.

Throughout this paper, we consider only those measurements that meet the conservative SOFIA/HAWC+ selection criteria: a polarization signal-to-noise ratio of $p/\sigma_p>3$; a polarization fraction of $p<50\%$; and a signal-to-noise ratio of the total intensity of $I/\sigma_I>200$ \citep{Gordon2018}.

\section{Magnetic Field in The Sgr C Complex}
\label{sec:bfield}

\subsection{Overview}
\begin{figure*}
    \centering
    \includegraphics[width=\textwidth]{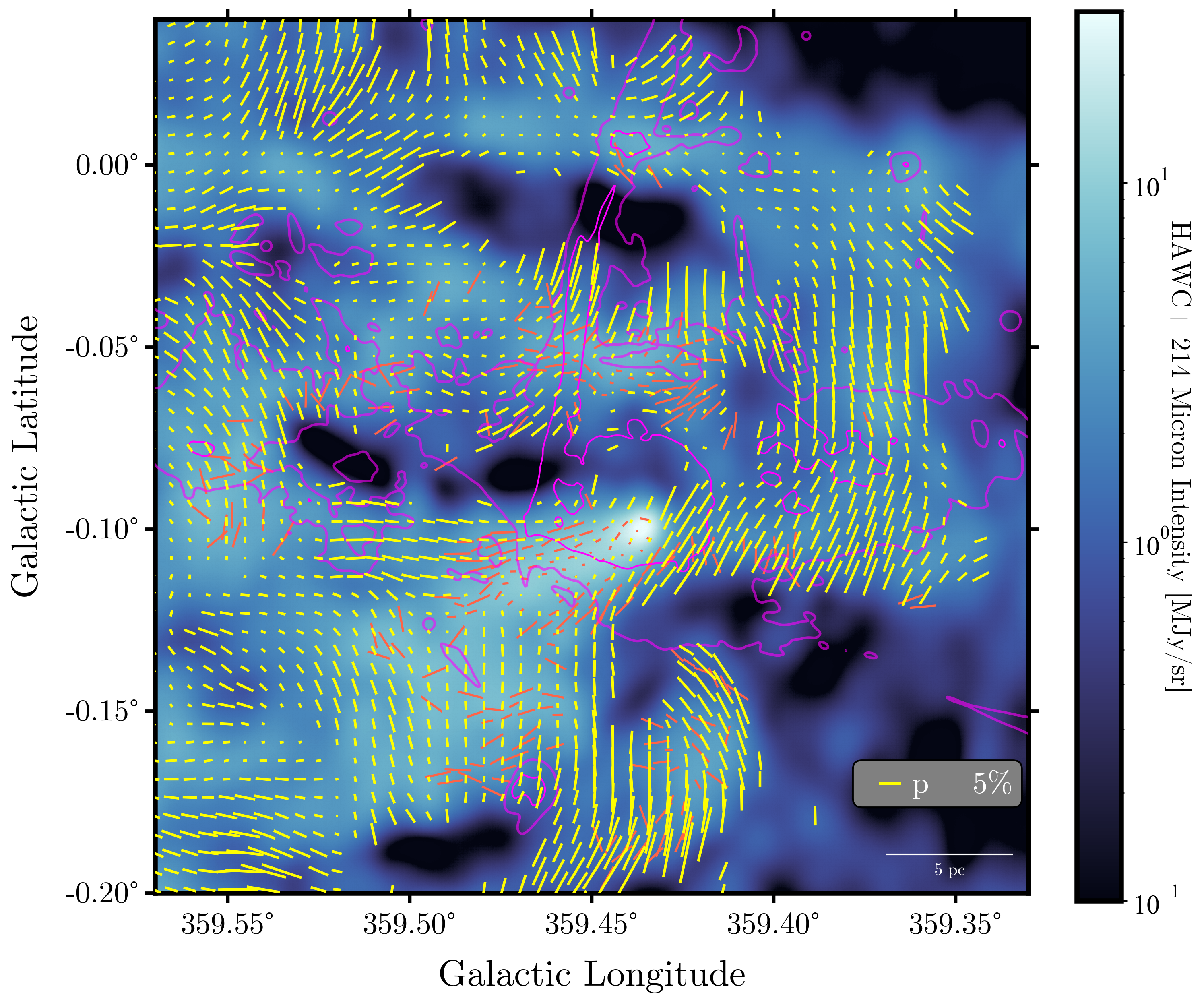}
    \caption{The FIREPLACE magnetic field pseudovectors in the Sgr C complex, displayed with the 214 \micron{} intensity image in the background, rendered using measurements of \citet{Pare2024}. The yellow dashes are the 214 \micron{} magnetic field pseudovectors made using Nyquist sampling at the HAWC+ beamsize, that fulfill the selection criteria described in Section~\ref{sec:observations}. The length of the vectors is scaled by their polarization fraction. The legend in the lower right corner denotes the length of a 5\% polarized pseudovector. The magenta contours are made from the MeerKAT/L-band observation of \citet{Heywood2022} at levels 0.0004 and 0.002 Jy/beam, with the brighter contour indicating higher intensity level. The orange dashes are the JCMT/POL2 magnetic field detections at 850 \micron{} \citep{Lu2024}. The JCMT/POL2 magnetic field measurement rendition share the same scaling relation with the FIREPLACE scale according to their polarization fraction. For visual clarity, we display all 850 \micron{} magnetic field detections with polarization fraction exceeding $p=0.08$ with a length equivalent to an 8\% polarized pseudovector.}
    \label{fig:214Bfield}
\end{figure*}

Figure~\ref{fig:214Bfield} presents the 214 \micron{} magnetic field measurements reported by \citet{Pare2024} as pseudovectors superposed on the Stokes I intensity image in the Sgr C region. 
A similar pseudovector representation of the magnetic field measurements is adopted throughout this paper. The 214 \micron{} emission is dominated by thermal emission from dust in molecular clouds, including the $-90$ \kms{} cloud in the north, the $-65$ \kms{} cloud in the south, and FIR-4. Throughout the extended region surrounding Sgr C, the 214 \micron{} polarization measurements that satisfy the selection criteria are clearly present with high significance throughout the far-IR-bright emission region. Measurements in faint regions do not commonly pass the selection criteria, owing either to low signal-to-noise ratio in either or both total intensity or polarization fractions, as outlined in Section~\ref{sec:observations}. Pseudovectors in faint regions can be contaminated by residual systematic effects and/or confused by unrelated emissions originating outside the CMZ (\citealt{Chuss2019, Butterfield2024}). However, in our field of view, our conservative selection criteria have largely mitigated such issues.

Throughout this paper, we compare our 214 \micron{} magnetic field orientations to the 850 \micron{} JCMT magnetic field orientations reported by \citet{Lu2024}, depicted as orange in Figure~\ref{fig:214Bfield}. The magnetic field measurements of \citet{Lu2024} were obtained with a spatial resolution similar to that in \citet{Pare2024} and were produced under a similar set of cutoff thresholds. From Figure~\ref{fig:214Bfield}, the 850 \micron{} magnetic field orientations do not generally agree with those at 214 \micron{}. It is likely that, given the large difference in the survey wavelengths, these two studies sample different regimes of dust temperature along the line of sight \citep[see, e.g., the dust temperature study reported by][]{Hankins2017}.

As was pointed out in \citet{Butterfield2024} and \citet{Pare2024} for other portions of the CMZ, the polarization fraction exhibits an anti-correlation with 214 \micron{} intensity. That is, the polarization fraction tends to be relatively higher in low-intensity regions and lower in high-intensity regions. This can either be due to the loss of grain alignment in denser regions, or due to the presence of multiple independent domains of different magnetic field orientations along the line of sight in the high-column density regions, therefore decreasing the integrated polarization fraction in high-column-density (and thus high-intensity) regions. The faint regions, typically present in the periphery of molecular clouds, appear more uniform in magnetic field orientation. The region where this anti-correlation is most obvious is the head-tail cloud, extending from the G359.44-0.10 EGO, where both 214 \micron{} and 850 \micron{} magnetic field detections exhibit extremely low polarization fractions, while the intensity is high at both of these wavelengths. 

In particular, we observe an elongated far-IR-dark region along $ b=-0.08$\degree{} in Figure~\ref{fig:214Bfield}. This region corresponds to the Galactic plane and is home to many compact mid-IR sources and short and curved radio filamentary structures (see Panel \#2 of Figure~\ref{fig:SgrC9}; \citealt{Liszt1995, Carey2009, Yusef-Zadeh2009, Hankins2020, Cotera&Hankins2024}). The low 214 \micron{} intensity and polarization non-detections there indicate a relative absence of cold dust, where it has either been heated by the local radiation field, expelled by stellar winds and expanding \hii{} regions, or there is simply a dearth of material there. 

\subsection{DCF Analysis}
\label{sec:DCF}
We adopt a modified DCF method \citep{Davis1951, Chandrasekhar1953} to calculate the sky-plane magnetic field strength using the polarization measurements from \citet{Pare2024}. The sky-plane magnetic field strength, following the DCF method, is expressed as 
\begin{equation}
    |\mathrm{\bf B}_{\rm POS}| = \sqrt{4\pi\rho_m}\frac{\sigma_v}{\sigma_\phi},
    \label{eq:DCF}
\end{equation}
where $\rho_m$ is the mass density, $\sigma_v$ is the velocity dispersion of the gas content, and $\sigma_\phi$ is the angular dispersion of the magnetic field pseudovectors. The DCF method is based on the assumption that the energy density of the turbulent or random gas motions is equal to the energy density of the fluctuations of the magnetic field about the local mean field, where its velocity dispersion is assumed to be isotropic. Though the general principle of the DCF method is commonly agreed upon, specific measurements of the input parameters of Equation~\ref{eq:DCF} are under active refinement and debate, especially in the context of a dynamically complex region like the GC (\citealt{Guerra2023, Butterfield2024ring, Lu2024}). In our DCF analysis, we do not consider the effect of shear on the sky-plane magnetic field strength $|\mathrm{\bf B}_{\rm POS}|$ (e.g. the shear-flow approximation considered by \citealt{Guerra2023}). Therefore, there is a caveat that our analysis can overestimate $|\mathrm{\bf B}_{\rm POS}|$ as the shear could lower the angular dispersion. However, due to the highly turbulent nature of the Sgr C region, shear does not play a significant role in the estimated magnetic field strength, as it is only significant in clouds with uniform motion. We address our methodology for estimating the inputs $\rho_m$, $\sigma_v$, and $\sigma_\phi$ to calculated the magnetic field strength $|\mathrm{\bf B}_{\rm POS}|$ as in Equation~\ref{eq:DCF} in the following subsections.

\subsubsection{Angular Dispersion ($\sigma_\phi$)}
\label{sec:ang_disp}
Past decades of research have shown that the GC magnetic field exhibits different behaviours viewed from different observational scales: a pervasive and ordered magnetic field in large-scale \citep{Novak2003, Planck2015, Mangilli2019, Guan2021}, a turbulent magnetic field driven by local environments in molecular clouds in small-scale \citep{Chuss2003, Butterfield2024, Butterfield2024ring, Lu2024, Pare2024b}, and a vertical inter-cloud field signified by the NTFs \citep{Serabyn1994}. However, in a dynamically complex region like the CMZ, the relative contribution of magnetic field systems on different scales is still under investigation. To this end, we follow the methods developed by \citet{Hildebrand2009} and \citet{Houde2009}, which have been implemented in numerous SOFIA/HAWC+ studies \citep{Chuss2019, Guerra2021, Butterfield2024, Butterfield2024ring}. We direct interested readers to \citet{Houde2009, Houde2016} for a detailed account of this method. The analysis in this subsection was implemented using the \textsc{polBpy} software \citep{PolBpy}.

We now detail the specific procedure, which separately accounts for the sky-plane magnetic field contribution from large and small scales. We write the sky-plane magnetic field as a superposition ${\bf B}_{\rm POS} = {\bf B}_0+{\bf B}_t$, where ${\bf B}_0$ is the ordered, large-scale field, and ${\bf B}_t$ is the turbulent, small-scale field. The angular dispersion can then be approximated as the turbulent-to-ordered magnetic field energy ratio \citep{Houde2009}:
\begin{equation}
    \sigma_\phi^2 \approx \frac{\langle {\bf B}_t^2\rangle}{\langle {\bf B}_0^2\rangle}.
    \label{eq:ang_disp}
\end{equation}

To determine this ratio, we adopt a two-component structure function, again separately accounting for the large- and small-scale magnetic field systems. As a statistical measure of the polarization angle difference $\Delta\phi(l)\equiv\phi(\mathbf{r+l})-\phi({\bf r})$ at a distance scale $l$, the structure function $1-\langle\cos{[\Delta\phi(l)]}\rangle$ can be computed using our polarization measurements. As such, fitting our two-component function to the empirical results (i.e. the FIREPLACE measurements) can constrain the contribution to the overall magnetic field strength from both scales, thereby yielding the required turbulent-to-ordered field ratio to calculate angular dispersion using Equation~\ref{eq:ang_disp}.  

The two-component structure function is written as \citep{Houde2009}
\begin{equation}
    1-\langle\cos{[\Delta\phi(l)]}\rangle = \frac{1-e^{-l^2/2(\delta^2+2W^2)}}{1+\mathcal{N}\left[\frac{\langle {\bf B}_t^2\rangle}{\langle {\bf B}_0^2\rangle}\right]^{-1}}+a_2l^2, 
    \label{eq:structurefunction}
\end{equation}
where $\delta$ is the correlation length for the turbulent field, $W$ is the beam radius, and $a_2$ is the large-scale contribution coefficient. Moreover, $\mathcal{N}$ is the number of turbulent cells along the line of sight defined by \citet{Houde2009} as 
\begin{equation}
\mathcal{N}= \frac{(\delta^2+2W^2)\Delta'}{\sqrt{2\pi}\delta^3},
\end{equation}
computed for a given region, where $\Delta'$ is the cloud's effective depth. In Equation~\ref{eq:structurefunction}, the first term on the right-hand side describes small-scale turbulence, and the second describes the large-scale contribution. 

\begin{figure}
    \centering
    \includegraphics[width=\columnwidth]{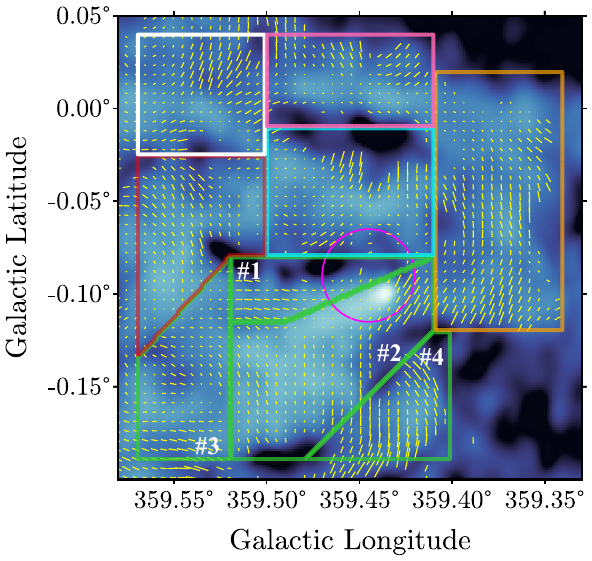}
    \caption{Individual regions used in the DCF analysis when fitting Equation~\ref{eq:structurefunction}, overlaid with the 214 \micron{} intensity image and pseudovector representation. The magenta circle marks the extent of the Sgr C \hii{} region. We note that the $-65$ \kms{} cloud is separated into four numbered regions (number-labelled green polygons).}
    \label{fig:masks}
\end{figure}

Before fitting Equation~\ref{eq:structurefunction}, we follow \citet{Houde2009} and infer $\Delta'$ using the shape of the autocorrelation function 
\begin{equation}
    \langle\bar{P}^2(l)\rangle \equiv \langle\bar{P}({\bf r})\bar{P}({\bf r}+{\bf l})\rangle,
    \label{eq:autocor}
\end{equation} 
where $\bar{P}({\bf r})$ is the beam-integrated polarized flux at coordinate ${\bf r}$. Instead of using the default array-by-array approach taken by the \textsc{polBpy} package, we use a Fast Fourier Transform to compute $\langle\bar{P}^2(l)\rangle$. Intuitively, Equation~\ref{eq:autocor} is a statistical measure of the average polarized flux within a distance $l$ in the sky plane, thereby giving the approximate spatial extent of the part of a cloud having the bulk of a coherent magnetic field orientation in the sky-plane. Assuming the cloud is isotropic, we posit that the distribution of polarized flux along the cloud's depth is similar to its sky-plane projection. We thereby follow \citet{Houde2009} and use the half-width at half-maximum (HWHM) value as the effective cloud depth. However, because the Sgr C complex consists of several regions identifiable by their spatial and radial velocity separation, or by the overall trends in their magnetic field orientations, the dispersion calculation only remains meaningful if applied to individual regions with their own identifiable magnetic field system. Therefore, we visually group the Sgr C magnetic fields based on their large-scale trends and only conduct DCF analyses using the magnetic field pseudovectors within a given region. The regions that we have thereby identified, along with their spatial extents, are shown in Figure~\ref{fig:masks}.
We list their computed effective cloud depths computed from Equation~\ref{eq:autocor} in Table~\ref{tab:clouddepth}.
\begin{table*}[]
    \centering
    \begin{tabular}{cccc}
    \hline\hline
    \textbf{Cloud} &  ~~\textbf{Colour/Number}~~ & ~~~~\textbf{$\Delta'$ [arcmin]}~~~~~  & \textbf{$\Delta'$ [pc]}  \\
    \hline
     $-65$ \kms{} cloud \#1 & Green \#1 & 0.64$\pm$0.15\arcmin{} & 1.45$\pm$0.36\\ 
     $-65$ \kms{} cloud \#2 & Green \#2 &  1.43$\pm$0.20\arcmin{}& 3.24$\pm$0.36\\
     $-65$ \kms{} cloud \#3 & Green \#3 & 1.03$\pm$0.08\arcmin{} & 2.45$\pm$0.19\\
     $-65$ \kms{} cloud \#4 & Green \#4 & 1.04$\pm$0.13\arcmin{} & 2.48$\pm$0.31\\
     FIR-4 & Pink & 0.84$\pm$0.13\arcmin{} & 2.00$\pm$0.31 \\
     Galactic plane & Red & 1.39$\pm$0.22\arcmin{} & 3.32$\pm$0.52 \\
     Source C& Orange & 1.41$\pm$0.14\arcmin{} & 3.36$\pm$0.33 \\
     Far Northeast & White & 1.36$\pm$0.07\arcmin{} & 3.24$\pm$0.17 \\
     $-90$ \kms{} cloud & Cyan & 1.09$\pm$0.12\arcmin{} & 2.60$\pm$0.29 \\
     \hline\hline
    \end{tabular}
    \caption{The effective cloud depths for individual regions shown in Figure~\ref{fig:masks}.}
    \label{tab:clouddepth}
\end{table*}
The individual clouds in Sgr C have a typical effective cloud depth of $\sim1$\arcmin{} or 2.38 pc. However, given the known line-of-sight overlap of clouds in Sgr C \citep{Lang2010}, the composite depths of this region could be much larger than interpreted on the level of individual clouds. More details regarding the autocorrelation function and effective cloud depth of each region can be found in Appendix~\ref{appx:A}.

\begin{figure*}
    \centering
    \includegraphics[width=\textwidth]{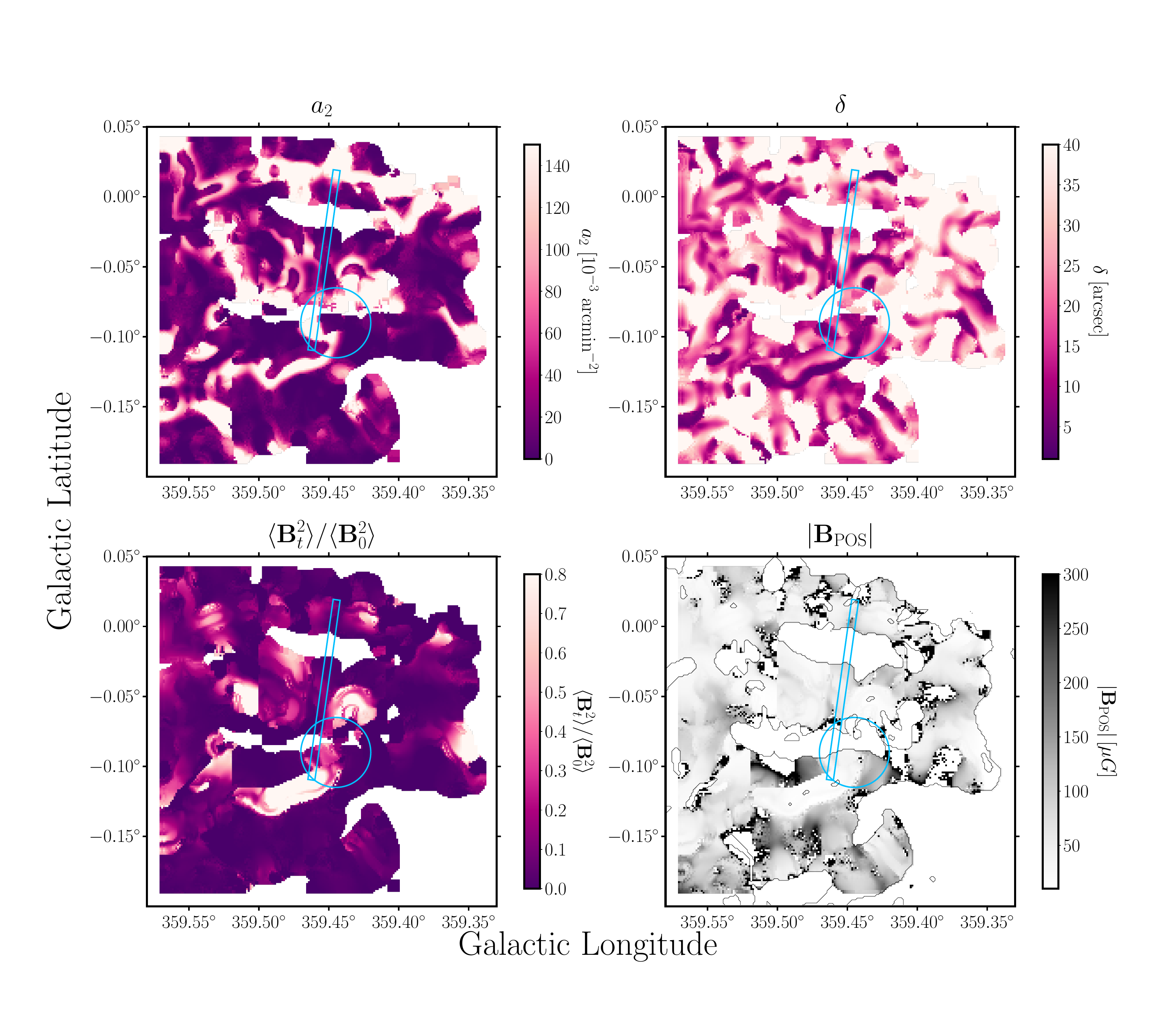}
    \caption{Images of the best-fit structure function variables $a_2$ (top left), $\delta$ (top right), $\langle {\bf B}_t^2\rangle/\langle {\bf B}_0^2\rangle$ (lower left), and the sky plane magnetic field strength $|{\bf B}_{\rm POS}|$ (lower right) in Sgr C. The blue circle and rectangle denote the approximate location of the Sgr C \hii{} region and NTF, respectively. We make a $2\sigma$ cut for the fitting variables prior to displaying these images. A small number of pixels with $|{\bf B}_{\rm POS}|>10~{\rm mG}$ was manually removed to eliminate unphysical field strength.}
    \label{fig:B_POS}
\end{figure*}

Having estimated the effective cloud depth, we perform a Markov Chain Monte Carlo (MCMC) fit of Equation~\ref{eq:structurefunction} for variables $\delta$, $a_2$, and most importantly, $\langle {\bf B}_t^2\rangle/\langle {\bf B}_0^2\rangle$ for all pixels in the field of view. The \textsc{polBpy} package uses the \textsc{emcee} package \citep{Foreman-Mackey2013} to generate MCMC posterior distributions of parameters $\delta$ and $a_2$, but do so for $\langle {\bf B}_t^2\rangle/\langle {\bf B}_0^2\rangle$ through a similar quantity $\Delta'(\langle {\bf B}_t^2\rangle/\langle {\bf B}_0^2\rangle)^{-1}$. All parameters are constrained to be positive with no upper bounds, though we do note that $\delta$ cannot be greater than the diameter of the dispersion function's circular kernel. The fit was performed on the structure function with $l\leq 1.25$\arcmin{} to ensure that only local dispersion is sampled. Here, we note that the behaviour of the kernel on the boundaries between regions can lead to improperly defined dispersion functions and therefore less-then-optimal fitted parameters. However, the MCMC fitting routine solves for the quantity $\Delta' (\langle {\bf B}_t^2\rangle/\langle {\bf B}_0^2\rangle)^{-1}$ for each pixel, which allows for using a pixel-dependent value of effective depth \citep[see][]{Guerra2021}. Furthermore, to improve the efficiency of the MCMC fits, we calculate and fit the dispersion functions using all available polarization measurements within one region shown in Figure~\ref{fig:masks}. The best-fit results are then used as the initial condition of the pixel-by-pixel MCMC fit to accelerate convergence. The final results were obtained with 300 iterations for each pixel, at which point we manually inspected the MCMC fit at numerous locations to ensure convergence. We found that the fit results do not evolve significantly after 150 iterations. The resulting images of the variables are shown in Figure~\ref{fig:B_POS}.

Globally, the $a_2$ and $\langle {\bf B}_t^2\rangle/\langle {\bf B}_0^2\rangle$ values are positively correlated, and they are negatively correlated with $\delta$. Given the high column density in many clouds in Sgr C, an increase in turbulence (which entails a decrease in turbulent correlation length) likely decreases the polarization fraction integrated along a line of sight. Therefore, the large-scale contribution ${\bf B}_0$ naturally becomes the dominant term relative to the turbulent contribution ${\bf B}_t$ in highly turbulent regions. We also note that, given the issue of moving kernel near regional boundaries, the $\langle {\bf B}_t^2\rangle/\langle {\bf B}_0^2\rangle$ results on the boundaries should be taken with caution. The observed discontinuities in Figure~\ref{fig:B_POS} are more likely due to unresolved $\delta$ parameter in some regions, which is a consequence of limited angular resolution.

Among the three high-density molecular clouds, we note that FIR-4 and the $-90$ \kms{} cloud exhibit more diverging magnetic field directions than the $-65$ \kms{} cloud, as can be observed in the cyan and pink regions in Figure~\ref{fig:masks}. In particular, the region immediately north of the Sgr C \hii{} region has a high turbulent-to-ordered ratio, which is likely induced by the compression from the ongoing expansion of the \hii{} region. On the contrary, the relatively uniform $-65$ \kms{} cloud and the source C cloud do not exhibit a significant spatial variation for these parameters, except for the area of the head-tail feature, where the polarization fraction is extremely low (see Figure~\ref{fig:214Bfield}). 

\subsubsection{Mass Density ($\rho_m$)}
\begin{figure}
    \centering
    \includegraphics[width=\columnwidth]{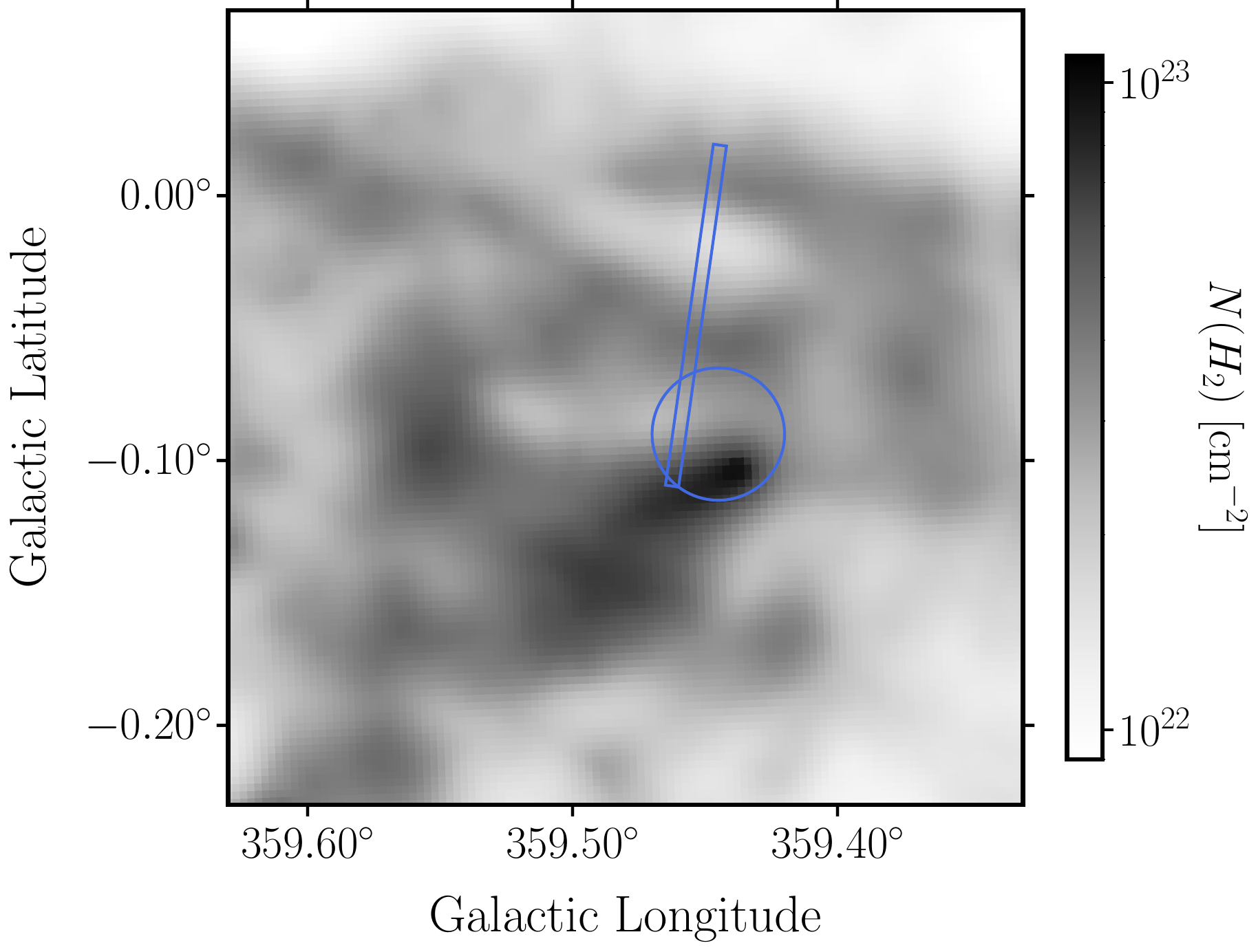}
    \caption{The column density map derived from fitting Equation~\ref{eq:I_nu} to the Hi-GAL 160, 250, and 350 \micron{} images, displayed in a logarithmic scale. The blue circle and rectangle indicate the locations of the Sgr C \hii{} region and the Sgr C NTF, respectively.}
    \label{fig:colden}
\end{figure}

We infer the mass density by deriving a column density image using the Hi-GAL survey data \citep{Molinari2010, Molinari2016}. Our procedure largely follows the spectral energy distribution (SED) fitting process described in section 3.1.3 of \citet{Chuss2019} and section 2.2 of \citet{Pare2024b}. Our SED fitting includes the Hi-GAL survey images from 160, 250, and 350 \micron{}. We did not include the 70 \micron{} image due to its uncertain extinction correction \citep{Compiegne2010, Battersby2011, Battersby2024}. The 500 \micron{} image was also excluded due to its low resolution. The images are convolved with a Gaussian kernel to match the 24\arcsec{} resolution of the SPIRE 350 \micron{} image. Following the convolution, we perform a pixel-by-pixel SED fitting to the single-temperature modified blackbody modelled by \citet{Vaillancourt2002}
\begin{equation}
    I_\nu(\nu, T) = (1-e^{-\tau(\nu)})B_\nu(T),
    \label{eq:I_nu}
\end{equation}
where $B_\nu$ is the Planck function. The optical depth $\tau(\nu)$ is defined as 
\begin{equation}
    \tau(\nu) = \epsilon \left(\frac{\nu}{\nu_0}\right)^\beta,
\end{equation}
where $\epsilon$ is a constant proportional to column density and ${\nu_0}$ is the frequency normalization constant adopted from \citet{Sadavoy2013}. For the fit, we leave $\epsilon$ and $T$ as free parameters while fixing $\beta=2$. Finally, the column density at each pixel is computed as
\begin{equation}
    N(H_2) = \frac{\epsilon}{\kappa_{\nu_0} \mu m_{\rm H}},
\end{equation}
where $\kappa_{\nu_0}$ is the reference dust opacity, $\mu$ is the mean molecular weight per hydrogen atom, and $m_{\rm H}$ is the atomic mass of hydrogen. We adopt ${\nu_0}=1000~{\rm GHz}$, $\kappa_{\nu_0}(1000~{\rm GHz}) = 0.1~ \mathrm{cm^2\,g^{-1}}$ and $\mu=2.8$ as the mean molecular weight of hydrogen. The derived column density map is shown in Figure~\ref{fig:colden}.
As expected, the cold molecular clouds, such as FIR-4, the $-65$ \kms{} cloud, and the $-90$ \kms{} cloud, which dominate the far-IR emission, are the most prominent in this map. We compare our column density result with that of \citet{Battersby2024}, which uses a slightly different method and also excludes the 70 \micron{} image, as well as \citet{Pare2024b}, which used the same method as this work but includes the 70 \micron{} Hi-GAL image. The column density values are similar across these works upon comparison. The mass density is then calculated in each region as 
\begin{equation}
    \rho_m = \mu m_H \frac{N(H_2)}{\Delta'},
\end{equation}
following \citet{PolBpy}.

\subsubsection{Velocity Dispersion ($\sigma_v$)} 
\label{sec:vel_disp}
Due to the highly turbulent nature of the clouds surrounding Sgr C, the velocity dispersion is quite large, as in most of the CMZ. We measure the velocity dispersion by taking the square root of the second spectral moment map of the HCN spectral cube provided by \citet{Jones2012} using the Common Astronomy Software Applications (\textsc{CASA}; \citealt{CASA}) package. Over the entire region, the velocity dispersion varies by only about 20\% around a mean value of $10$ \kms{}, and the mean value within each of our defined regions varies by even less. Meanwhile, there are many instrumental artifacts contained in the velocity dispersion map calculated from Mopra spectral cube, which cannot be easily mitigated. Given the inherent uncertainty in our sky-plane magnetic field strength determinations owing to the fact that we only have access to the intensity-weighted average of the velocity dispersion along the line of sight, and similarly for the magnetic field orientations, we have chosen to adopt a constant velocity dispersion of $10$ \kms{}. The resulting field strength estimates vary linearly with the adopted velocity dispersion (c.f., Equation~\ref{eq:DCF}). 

\subsubsection{Magnetic Field Strength Results}
\label{sec:B_POS}
The sky-plane component of the magnetic field strength calculated from the above-described inputs using Equation~\ref{eq:structurefunction} is displayed in the lower right panel of Figure~\ref{fig:B_POS}. The field strength exhibits a large range from $\sim 30~\mu {\rm G}$ to $\sim300~{\rm \mu G}$. This range agrees with many past attempts to infer the CMZ magnetic field strength \citep{Crocker+10a, Chuss2003, Kruijssen2014, Sormani2020, Tress2024}. We find that the morphology of the angular or velocity dispersion image has a limited effect on the inferred $|{\bf B}_{\rm POS}|$, but the local turbulence of a region is anti-correlated with $|{\bf B}_{\rm POS}|$ (shown in the bottom panels of Figure~\ref{fig:B_POS}). Therefore, the dominant factor affecting $|{\bf B}_{\rm POS}|$ is the local shear and compression. On the scale of individual molecular clouds, this conclusion suggests that the most turbulent clouds (FIR-4, the $-90$ \kms{} cloud, and the $-65$ \kms{} cloud) exhibit the lowest $|{\bf B}_{\rm POS}|\sim 100~\mu {\rm G}$. 

The head-tail cloud (which extends from the southern portion of the Sgr C \hii{} region, see Figure~\ref{fig:SgrC_finding}) displays intriguing features from our DCF analysis. The low polarization fraction ``tail" detected in both 214 and 850 \micron{} surveys extending from the G359.45-0.10 ``head" displays a consistently lower $|{\bf B}_{\rm POS}|$, creating a sharp contrast with the high field strength regions in clouds with uniform field direction. Once more, the low $|{\bf B}_{\rm POS}|$ in the head-tail cloud likely signifies its high turbulence and complex composition (i.e. several clouds likely superposed along the same line of sight due to the high column density and low polarization fraction detected in both 214 \micron{} and 850 \micron{}). 

We also highlight the morphology of the field near the Sgr C \hii{} region. In the north of the \hii{} region around $l=359.43$\degree{}, $b=-0.07$\degree{}, a half-shell of low field strength is apparent. Simultaneously as a high turbulent-to-ordered ratio region, we speculate that the ionization front of the \hii{} region may be the reason behind the high turbulence. Although the molecular cloud there has a high column density, the turbulence induced by the shock and ionization front driving the expansion presumably leads to an overall low $|{\bf B}_{\rm POS}|$.

We note a few caveats in our DCF analysis. One should be cautious that the FIREPLACE survey is conducted in a single wavelength. The 214 \micron{} filter tends to sample clouds with colder temperatures (colder compared to the warmer Sgr C \hii{} region; \citealt{Carey2009}, \citealt{Hankins2020}), which in some regions may only be one of multiple clouds along the line of sight. The most obvious example in Sgr C is a cloud centred at $l=359.48$\degree{}, $b=-0.17$\degree{} with a broad range of velocities (from -92.6 \kms{} to -34.3 \kms{}) detected in the HCN line analysis presented in Section~\ref{sec:spectral} (see Figure~\ref{fig:HCN_vel}). The 850 \micron{} measurements from \citet{Lu2024} nicely trace the morphology of this cloud but disagree with the 214 \micron{} magnetic field orientations in this region, which may indicate that the 850 \micron{} measurements are sampling a different cloud along that line of sight, which may naturally possess a different magnetic field orientation. As such, inferring $|{\bf B}_{\rm POS}|$ using only one wavelength introduces some uncertainty, since different clouds, or even the same cloud with different temperature domains, along a given line of sight can have independent field directions that then lead to a reduction of the polarization fraction. Another possible source of error comes from the column density estimation, as we assume a single-temperature blackbody, having multiple clouds (possibly with different temperatures) along the same line of sight could cause an error in our column density estimation, which also affects the $|{\bf B}_{\rm POS}|$ calculation. In regions dominated by higher temperatures, the DCF analysis cannot be reliably performed since there is an absence of valid 214 \micron{} magnetic field detections.

\section{Multi-wavelength Observations}
\label{sec:discussion}
This section provides an overview of the relevant features of Sgr C by comparing existing multi-wavelength studies with the FIREPLACE 214 \micron{} magnetic field measurements.

\subsection{Assembly of Observations}
\begin{figure*}
    \centering
    \includegraphics[width=\textwidth]{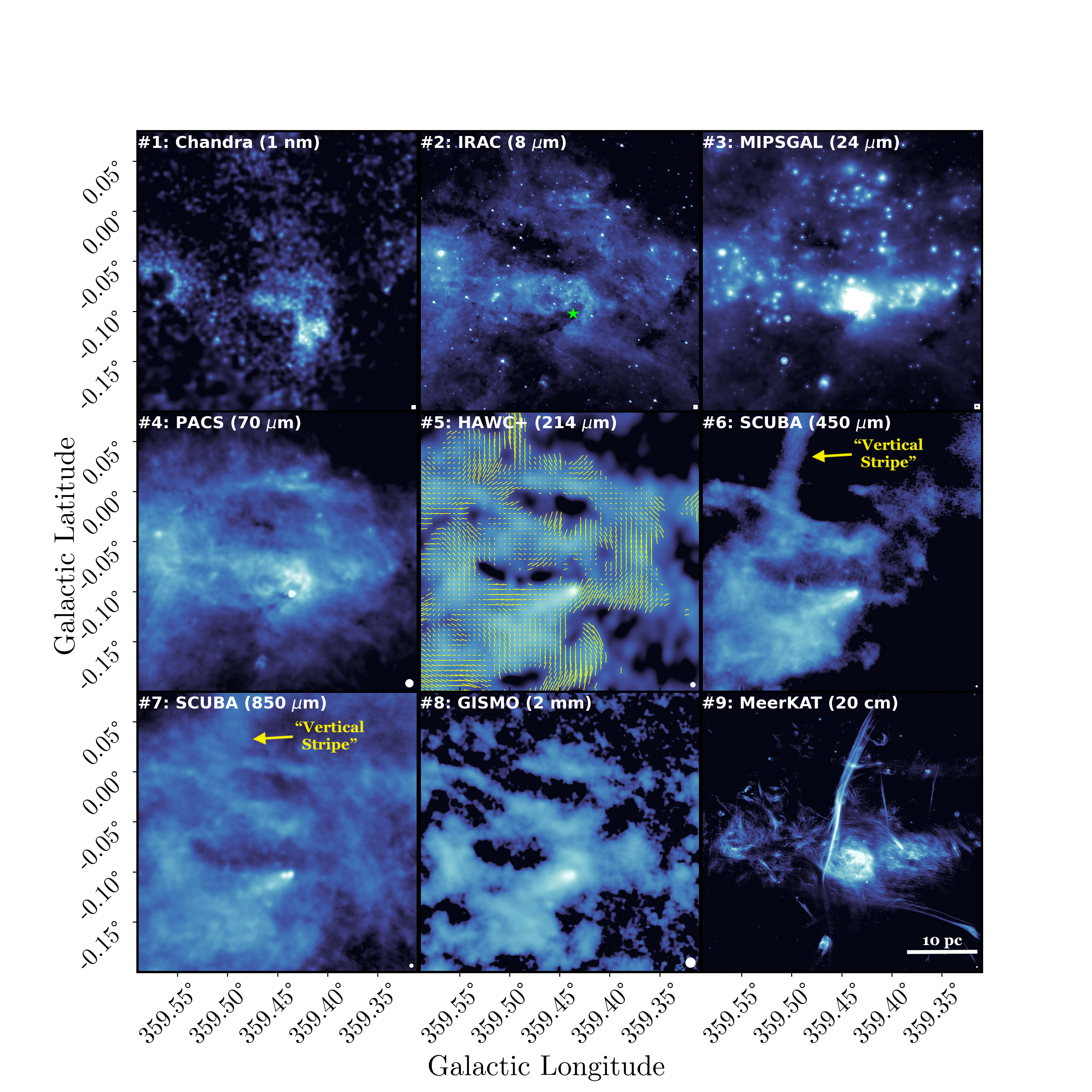}
    \caption{Selected multi-wavelength images of the Sgr C complex. The region has been surveyed in $1-4~{\rm keV}$ X-rays by {\it Chandra} \citep{Wang2021}, 8 \micron{} by {\it Spitzer}/IRAC \citep{Stolovy2006}, 24 \micron{} by {\it Spitzer}/MIPSGAL \citep{Carey2009}, 70 \micron{} by {\it Herschel}/PACS \citep{Molinari2010}, 214 \micron{} by SOFIA/HAWC+ \citep{Pare2024}, 450 \micron{} and 850 \micron{} by JCMT/SCUBA \citep{Pierce-Price2000}, 2 mm by IRAM/GISMO \citep{Arendt2019}, and at 20 cm by MeerKAT \citep{Heywood2022}. For visual comparison between the multi-wavelength images and the magnetic field detections, we add the 214 \micron{} magnetic field pseudovectors from \citet{Pare2024} in Panel \#5 as yellow segments. The corresponding beam sizes are displayed as white circles in the lower right corner of each image. The Green star in Panel \#2 labels the G359.44-0.10 EGO, and the yellow arrows indicate in Panels \#6 and \#7 indicates the location of a vertical structure in these maps. All images are displayed with a logarithmic stretch. }
    \label{fig:SgrC9}
\end{figure*}

Figure~\ref{fig:SgrC9} shows nine intensity images covering a broad range of wavelengths, from 1 nm to 20 cm, to highlight the significant structures in the Sgr C region. 
Of the published surveys of the Sgr C region, the images presented in Figure~\ref{fig:SgrC9} have been chosen to avoid repetition. For example, the 870 \micron{} APEX/ATLASGAL survey \citep{Schuller2009} is not included because it exhibits features nearly identical to those in the JCMT/SCUBA 850 \micron{} survey \citep{Pierce-Price2000}. 

The Galactic plane appears bright in the IR regime out to 70 \micron{} and in radio, with the peak emission at these wavelengths arising near the Sgr C \hii{} region. In mid-IR and radio images (Panels \#2--4, and \#7 in Figure~\ref{fig:SgrC9}), the Sgr C \hii{} region exhibits a similar elliptical morphology elongated along the Galactic plane. We observe an absorption ``tail" originating from inside the Sgr C \hii{} region extending towards the southeast, where the EGO is located (G359.44-0.10; \citealt{Kendrew2013, Lu2019, Lu2021, Crowe2024}; labelled in Panel \#2 of Figure~\ref{fig:SgrC9}). This low-intensity ``valley" is manifested at 8, 24, and 70 \micron{} and is a result of foreground absorption by the head-tail cloud, which turns to emission at far-IR wavelengths beyond 70 \micron{}. The mid-IR intensity images tend to become more diffuse at longer wavelengths, which is partially due to the trend that lower-temperature dust is generally the most extended, and partially due to decreased resolution at longer wavelengths. 

The 8 \micron{} emission traces the polycyclic aromatic hydrocarbons (PAHs) ionized by stellar UV radiation \citep{Stolovy2006, Narayanan2023}, thereby showing the extent of the elliptical region impacted by UV emission and star formation around the Sgr C \hii{} region. In the 8 to 70 \micron{} interval (Panels \#2--4 in Figure~\ref{fig:SgrC9}), the emission also traces warm dust heated by stellar emission. In this band, one could also observe the ``head" of the head-tail structure as an extremely compact and luminous source (classified as an EGO by \citealt{Kendrew2013}) to the immediate south of the \hii{} region, also becoming more diffuse towards longer wavelengths. 

In the far-IR regime (214--850 \micron{}, Panels \#5--7 in Figure~\ref{fig:SgrC9}), the images are dominated by thermal emission from cold dust residing in molecular clouds. The Galactic plane and the \hii{} region, by contrast, appear dark in the far-IR regime. The ``head" of the head-tail structure is larger in size at far-IR wavelengths than in the mid-IR. Though this increase in size could be a result of cold dust being more diffuse, it can again be a result of lower spatial resolution at far-IR wavelengths. The ``tail" also appears bright at these wavelengths, reflecting a relatively high column density of cold dust. Indeed, we have an inferred column density of $\sim10^{23}~\mathrm{cm^{-2}}$ in this region, shown in Figure~\ref{fig:colden}. 

A noteworthy structure is the vertical ``stripe" extending northwards from $l=359.52$\degree{}, $b=-0.05$\degree{} that appears in the 450 and 850 \micron{} SCUBA images ( labelled by orange arrows in Panels \#6--7 in Figure~\ref{fig:SgrC9}), though it is most noticeable at 450 \micron{} and is much more diffuse at 850 \micron{}. With the caveat that this could be an artifact due to filtering, we note that the northern end of this ``stripe" coincides with the mid-IR ridge that connects the extended thermal source AFGL5376 and the Galactic plane \citep{Uchida1990, Uchida1994}. The 214 \micron{} magnetic field orientations do not bear any obvious relationship to the morphology of this stripe. We go no further than noting the above correspondence, as a discussion of AFGL5376 is beyond the scope of this paper. 

As a transition between far-IR and radio wavelengths, the GISMO image at 2 mm displays characteristics from both regimes, showing a combination of thermal emission from cold dust and free-free synchrotron emission from ionized gas. As such, we see not only the 100 pc ring dominated by cold dust, but also the ionized gas in the Sgr C \hii{} region. 

The radio continuum emission (Panel \#9 in Figure~\ref{fig:SgrC9}) is produced by free-free emission, primarily from the Sgr C \hii{} region and by synchrotron emission from the nearby NTFs \citep{Heywood2022, Yusef-Zadeh2022index, Yusef-Zadeh2022}, the most prominent of which is the Sgr C NTF bundle that appears to originate near the Sgr C \hii{} region. Although the peak emission from the Sgr C NTF is at $b\gtrsim-0.10$\degree{} and is most obviously seen to extend towards the north, there is a faint extension towards the south which ends near the relatively compact elliptical or bipolar radio source G359.46-0.17. We hypothesize that the abrupt intensity change in the NTF is caused by its interaction at that point with the high-density gas in the head-tail structure (detailed in Section~\ref{sec:NTF}). At that location, the diffusion of relativistic electrons to the south along the field lines is largely, but apparently not entirely, blocked by the presence of the dense gas. Furthermore, if the interaction with the cloud is taking place along the ionization front of the Sgr C \hii{} region, the acceleration of the electrons to relativistic energies could be the result of magnetic field line reconnection there, as proposed by \citet{Serabyn1994} or by diffusive shock acceleration if a strong stellar wind creates a shock front there \citep{Rosner1996}. 

There are also abundant short and curved radio filamentary structures to the east of the Sgr C \hii{} region along the Galactic plane, as noted in \citet{Liszt1995}. Some of these radio structures can be non-thermal, as inferred by the absence of IR counterparts. Their relative faintness could simply be interpreted in terms of a smaller population of relativistic electrons. We also do not observe any correspondence between the FIREPLACE magnetic field orientations and these short radio filaments.

\subsection{Spectral Line Analyses}
\label{sec:spectral}
To better understand the dynamics of Sgr C, we require the relative position and motion of the observed structures. Here, we appeal to the 3 mm Mopra spectroscopic survey conducted by \citet{Jones2012} and the \hi{} absorption analysis by \citet{Lang2010} for velocity constraints and line-of-sight placements. 

\subsubsection{Structure of Clouds in Sgr C}
\label{sec:HCN}
\begin{figure*}[p]
    \centering
    \includegraphics[width=\textwidth]{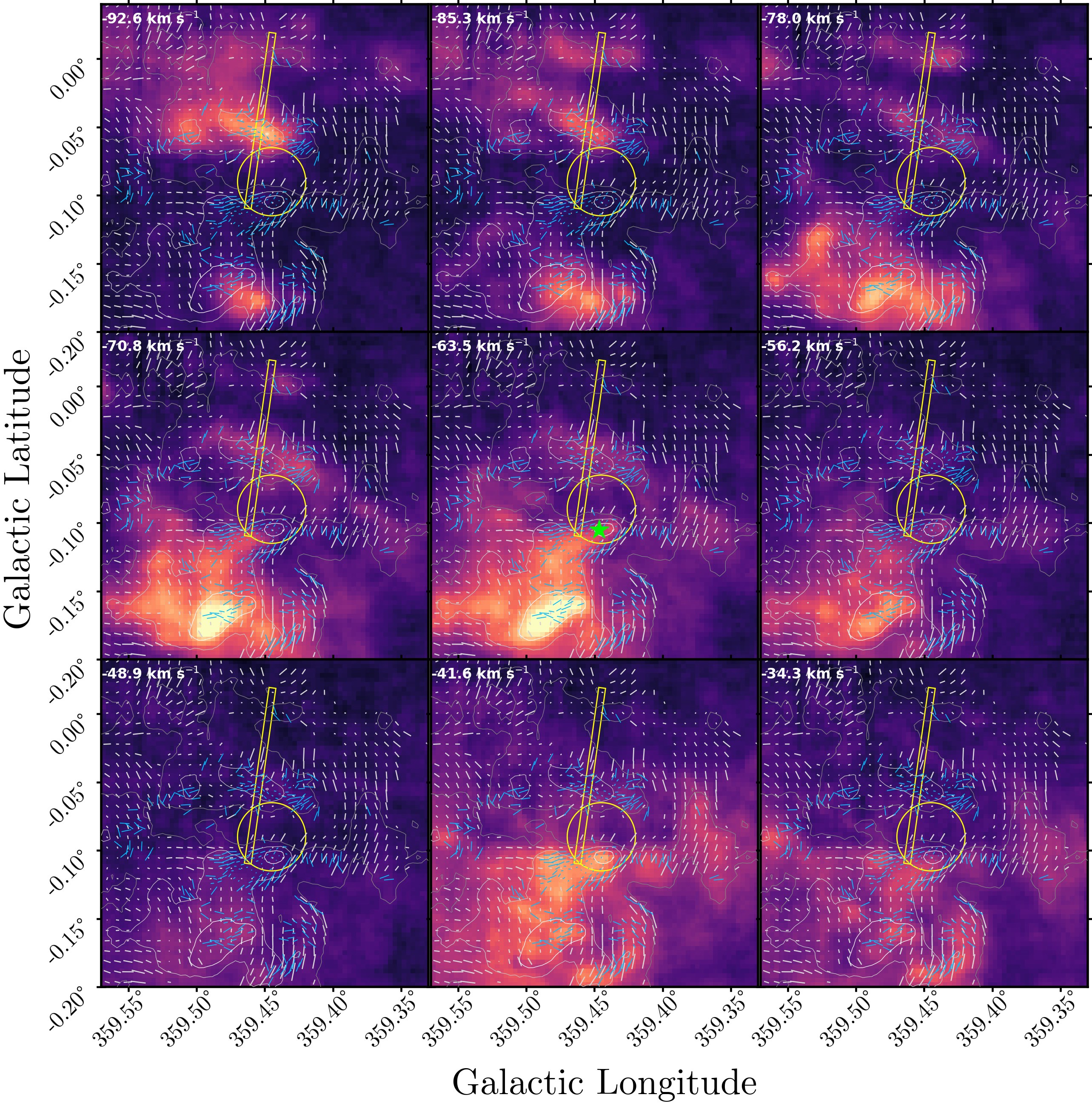}
    \caption{Velocity channel maps of the HCN J=1-0 line measured by the 3 mm Mopra survey \citep{Jones2012}. The survey has spatial and spectral resolutions of 36\arcsec{} and 1.84 \kms, respectively. The mean velocity of each channel is indicated in the top left corner. These channel maps cover the velocity range between $-92.6$ and $-34.3$ \kms{} in steps of $7.3$ \kms{}. The grey-scale contours show the peak emission of HCN integrated through all observed velocities drawn at 0.75, 1.25, 1.75 K \kms{}, with the brighter contours indicating higher intensity. The white line segments are the 214 \micron{} FIREPLACE magnetic field pseudovectors from \citet{Pare2024} and the blue segments are the 850 \micron{} magnetic field pseudovectors from \citet{Lu2024}. The yellow schematics indicate the approximate placement and extent of the Sgr C \hii{} region and the Sgr C NTF. The green star in the central panel denotes the location of the G359.44-0.10 EGO, which is addressed in detail in Section~\ref{sec:headtail}. The global low intensity in channels near $-50$ \kms{} is due to absorption in the foreground ``3 kpc arm" \citep{Lang2010}. The 214 \micron{} magnetic field pseudovectors in this figure are rendered using Nyquist sampling at the Mopra beam size. }
    \label{fig:HCN_vel}
\end{figure*}

We select the HCN J=1-0 line to deduce the dynamical structure of the clouds in Sgr C to aid our understanding of the FIREPLACE magnetic field orientations. The HCN J=1-0 line is one of the most prominent lines in the GC, arising in relatively dense gas \citep{Jones2012}. Figure~\ref{fig:HCN_vel} shows the velocity channel maps of this line in the velocity range from $-92.6$ \kms{} to $-34.3$ \kms{}. 
We note that a foreground cloud often referred to as the ``3 kpc arm" causes significant absorption around $-50$ \kms{} \citep{Lang2010}. The channel maps in Figure~\ref{fig:HCN_vel} show a coherently decreasing radial velocity from the north of the Sgr C \hii{} region ($\gtrsim -90$ \kms, which is constituted of FIR-4 and the $-90$ \kms{} cloud) to the southeast ($\sim -65$ \kms{}, which is constituted of the head-tail cloud and the $-65$ \kms{} cloud). These velocities follow the 100 pc dust ring revealed by \citet{Molinari2011}. The radial velocity of FIR-4 (G359.45+0.02) is similar to that of the nearby $-90$ \kms{} cloud, which is a part of the 100 pc ring. FIR-4 is therefore likely physically related to the 100 pc ring.

The Sgr C \hii{} region is largely devoid of HCN emission, except for the intruding head-tail cloud. Both the radio recombination line measurements reported by \citet{Liszt1995} and \hi{} absorption analysis by \citet{Lang2010} determined the velocity of the \hii{} region to be $\sim-60$ \kms. Given the velocity structure observed in Figure~\ref{fig:HCN_vel}, we infer that the $-90$ \kms{} cloud, the $-65$ \kms{} cloud, and the Sgr C \hii{} region constitute a continuous structure linked to the 100 pc ring, and are likely intrinsically related. We direct the reader to more details of the line-of-sight placement of structures in Sgr C to Section 3.6.3 of \citet{Lang2010}.

In the southern portion of the $-65$ \kms{} cloud, there is a bright bulk of HCN emission at $l=359.47$\degree{}, $b=-0.17$\degree{} visible at all displayed velocities. In this region, we do not find any morphological relationship between the HCN emission and the 214 \micron{} magnetic field orientations. However, many 850 \micron{} magnetic field detections were reported in this cloud by \citet{Lu2024}, and they do not generally agree with the 214 \micron{} results of \citet{Pare2024}. The HCN emission offers several possible explanations for this discrepancy. It could be that this is a particularly disturbed and turbulent cloud with multiple temperature components, or there could be multiple clouds superposed along the same line of sight, as we previously suggested in Section~\ref{sec:B_POS}. Both can cause the magnetic field measured from different wavelengths to exhibit different orientations. It is also likely that this cloud has a radial velocity gradient along the line of sight, given that this is where the rotating 100 pc ring reverses direction in projection. The observation of this particularly broad-linewidth cloud re-affirms our caveat for the DCF analysis in Section~\ref{sec:B_POS} regarding the incomplete accounting for multiple line-of-sight clouds having different velocities in the derivation of the magnetic field strengths.

The previously mentioned 214 \micron{} magnetic field measurement cavity at $l=359.44$\degree{}, $b=-0.13$\degree{} between the $-65$ \kms{} cloud and Cloud \#4 (centred at $l=359.43$\degree{}, $b=-0.17$\degree{}; see Figure~\ref{fig:masks}) is of interest to be studied. In Figure~\ref{fig:HCN_vel}, the $-65$ \kms{} cloud and Cloud \#4 appear to be portions of a continuous structure so they are likely physically related. We hypothesize that they started as one cloud and were later separated into the currently observed shape by an outside force that created the magnetic field measurement cavity at $l=359.44$\degree{}, $b=-0.13$\degree{}. We note a potentially relevant X-ray association with this cavity in Section~\ref{sec:X-ray}.

\subsubsection{Shock Tracer in the Twisted 100 pc Ring}
\begin{figure*}
    \centering
    \includegraphics[width=\textwidth]{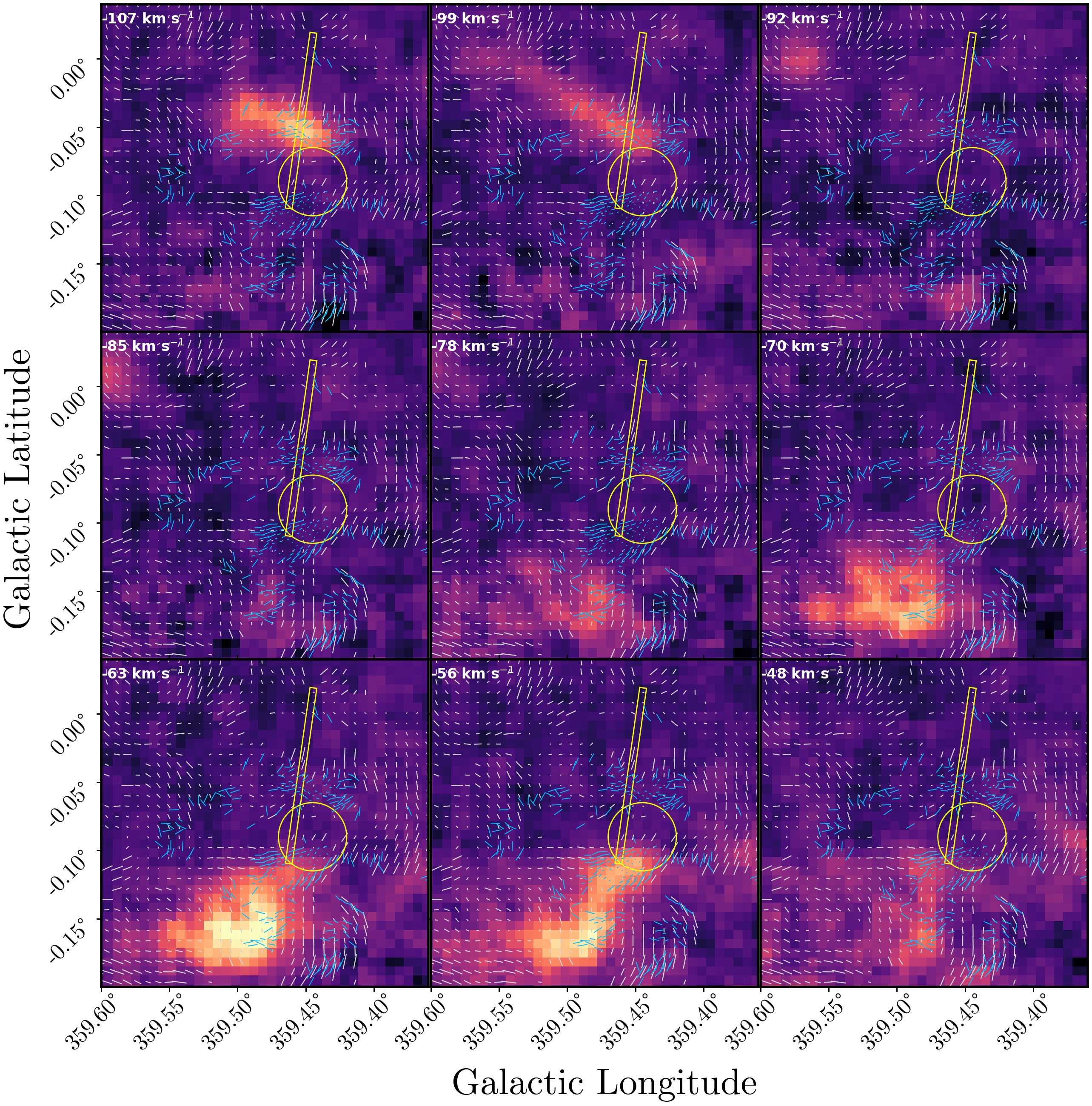}
    \caption{Velocity channel maps of SiO J=2-1 emission between $-110$ and $-49$ \kms, measured by the Mopra survey \citep{Jones2012}. The 214 \micron{} and 850 \micron{} magnetic field pseudovectors are superimposed as white and blue line segments, as in Figure~\ref{fig:HCN_vel}. The yellow schematic indicates the approximate location and extent of the Sgr C \hii{} region and the Sgr C NTF.}
    \label{fig:SiO_vel}
\end{figure*}

As mentioned in the introduction, Sgr C is located on the western vertex of a dust ring \citep{Molinari2011}, and is postulated to be a collision site of $X_1$-$X_2$ orbits due to the bar potential \citep{Sormani2018, Sormani2020}, which likely induces shocks and star formation. Meanwhile, the expanding Sgr C \hii{} region can also induce a compression front that collides with surrounding molecular clouds \citep{Liszt1995} and add to the effect of the large-scale shocks. We employ the many 214 \micron{} magnetic field detections toward the molecular clouds in the Sgr C region to examine how such shocks might be manifested in the magnetic field orientations. Figure~\ref{fig:SiO_vel} shows the velocity channel maps of the SiO J=2-1 line, a known shock tracer also from the Mopra survey \citep{Jones2012}.
The SiO channel maps exhibit a similar velocity structure to that of HCN, indicating that the SiO emissions are from within the 100 pc ring. However, the SiO emission appears much less spatially extended than that of HCN and more concentrated around the Sgr C \hii{} region. This local enhancement of SiO indicates that the portions of the 100 pc ring surrounding the Sgr C \hii{} region are particularly subjected to shocks, likely due to a combination of 1) the impact of the dust lane inflowing along the Galactic bar upon the CMZ material in the 100 pc ring, and 2) the collision between ambient gas and the expanding shell surrounding the \hii{} region \citep{Riquelme2025}. The collision likely increases the turbulence within the 100 pc ring, ultimately leading to a low polarization fraction, a higher turbulent-to-ordered magnetic field strength ratio, and relatively low inferred sky-plane magnetic field strength, as shown in Figure~\ref{fig:B_POS}.

\section{Individual Features in the Sgr C Complex}
\label{sec:features}
In this section, we use the 214 \micron{} magnetic field measurements from \citet{Pare2024} along with multi-wavelength images to draw inferences about individual features in the Sgr C complex. 

\subsection{Associated NTFs \& Radio Features}
NTFs are valuable probes of the large-scale magnetic field in the inter-cloud volume of the central few hundred parsecs of the Galaxy \citep{Serabyn1994, Morris2006, Thomas+20}. The following subsections examine the magnetic field behaviour around the NTFs and radio structures observed in the Sgr C region, keeping in mind that the FIREPLACE detections primarily probe the magnetic field within clouds.

\subsubsection{The Sgr C NTF}
\label{sec:NTF}
\begin{figure*}
    \centering
    \includegraphics[width=\textwidth]{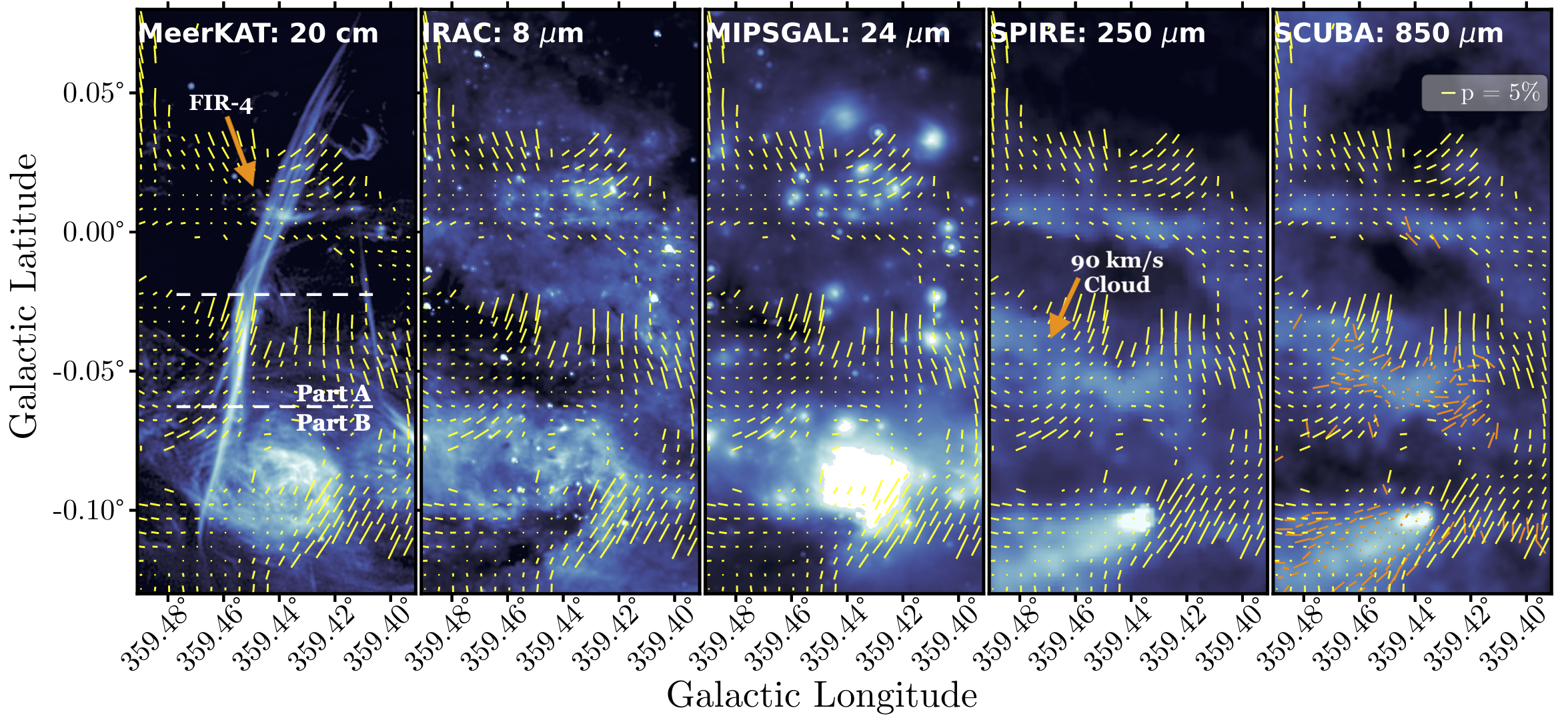}
    \caption{The 214 \micron{} magnetic field pseudovectors near the Sgr C NTF and G359.43+0.02 YSO superimposed on emission at 20 cm \citep{Heywood2022}, 8 \micron{} \citep{Stolovy2006}, 24 \micron{} \citep{Carey2009}, 250 \micron{} \citep{Molinari2010}, and 850 \micron{} \citep{Pierce-Price2000}. The yellow segments are the 214 \micron{} magnetic field pseudovectors from \citet{Pare2024} and the orange segments are the 850 \micron{} magnetic field pseudovectors from \citet{Lu2024}. Important features in this region, such as FIR-4, the two parts of the Sgr C NTF, and the 90 \kms{} cloud, are labelled by orange arrows in different panels where the mentioned structures are visible.} 
    \label{fig:NTF}
\end{figure*}

Figure~\ref{fig:NTF} shows the FIREPLACE 214 \micron{} magnetic field pseudovectors superimposed upon five selected images covering the Sgr C NTF. 
This extended NTF appears luminous and compact below $l=-0.02$\degree{} and it splits into two parallel filaments and becomes fainter and more diffuse above this latitude. This NTF overlaps on the sky-plane with three distinct magnetic field intervals:
\begin{itemize}
    \item{\it(i)} Below $b=-0.10$\degree{}, where the faint end of this NTF disappears in the head-tail cloud and the $-65$ \kms{} cloud (detailed in Section~\ref{sec:radioriver})
    \item{\it (ii)} $-0.08$\degree{}$\lesssim b\lesssim-0.02$\degree{}, where the FIREPLACE magnetic field detections arise from the $-90$ \kms{} cloud.
    \item{\it (iii)} Above $b=0$\degree{}, where this NTF overlaps with a magnetic field system dominated by FIR-4 (detailed in Section~\ref{sec:FIR-4}).
\end{itemize}
Interval \textit{(ii)} above is roughly the extent of Part A as defined in \citet{Roy2003} and \citet{Lang2010} (also labelled in the leftmost panel of Figure~\ref{fig:NTF}). We note additionally that the Sgr C NTF has a faint end towards the south, detailed in Section~\ref{sec:radioriver}.

In the radio continuum below $l=-0.02$\degree{} (indicated by the upper dashed line in the leftmost panel of Figure~\ref{fig:NTF}), the NTF has the same width but Part A is more luminous than Part B (the part of the Sgr C NTF below $l=-0.08$\degree{} labelled in Figure~\ref{fig:NTF}; \citealt{Roy2003}). The radio spectral index analysis of \citet{Yusef-Zadeh2022index} indicates that Part A has an average spectral index of $\sim-1.02$ and Part B has an average spectral index of $\sim-0.65$. In Part A, the magnetic field orientation near the NTF roughly follows its direction. As our magnetic field measurements sample the field from relatively cold clouds, the direction of the FIREPLACE magnetic field in the $-90$ \kms{} cloud (labelled in the fourth panel in Figure~\ref{fig:NTF}) is consistent with the orientation of the Sgr C NTF, as are the 850 \micron{} measurements of \citet{Lu2024}. Moreover, the DCF analysis from Figure~\ref{fig:B_POS} shows that there is a ``tunnel" of low turbulent-to-ordered ratio and high $|{\bf B_{\rm POS}}|$ along the Sgr C NTF. However, such alignment ceases beyond $l>-0.02$\degree{} due to an absence of magnetic field detections outside the molecular cloud. Additionally, the NTF becomes much more diffuse after departing from the molecular cloud toward the north. Above $l=0$\degree{}, the magnetic field orientation and the Sgr C NTF are no longer aligned, which could be attributed to strong far-IR emission from FIR-4 dominating the magnetic field detections in that region. 

\subsubsection{FIR-4: G359.43+0.02}
\label{sec:FIR-4}
At the northern portion of the Sgr C NTF where it splits into two filaments and begins to fade out toward the north, lies a short, linear radio feature oriented parallel to the Galactic plane along $b\approx 0.01$\degree{}, labelled in the leftmost panel of Figure~\ref{fig:NTF}). It was originally found and denoted as FIR-4 by \citet{Odenwald1984} in the course of their far-IR survey of the GC. FIR-4 and its counterparts are displayed in Figure~\ref{fig:NTF}. The mid-IR counterpart of FIR-4 is the G359.43+0.02 YSO Cluster reported by \citet{Yusef-Zadeh2009} and prominently displayed in the central panel of Figure~\ref{fig:NTF}.

Given its moderately strong far-IR brightness, many significant 214 \micron{} magnetic field detections were made in FIR-4. Overall, the magnetic field exhibits a divergent morphology with the magnetic field pointing in the radial direction away from the centre of the source. The polarization fraction in the centre and near the radio thread is mostly small, but increases towards the periphery. The $|{\bf B}_{\rm POS}|$ in the centre is also lower than that in the periphery (c.f., Figure~\ref{fig:B_POS}). We hypothesize that this morphology likely traces the outflow driven by protostellar winds, which induce high turbulence within the cluster but a collective, ordered outflow on the periphery. In particular, the radio counterpart of FIR-4 coincides with the region along $l=0.01$\degree{} with the lowest polarization fraction. The radio and mid-IR emission from FIR-4, arrayed as they are along the base of the embedded stellar cluster, suggest that UV emission from nearby YSOs has heated the local dust and has produced an ionization front, which appears linear in projection. 

FIR-4 intersects with the Sgr C NTF in projection on the plane of the sky. \citet{Lang2010} suggested that FIR-4 could have the same line-of-sight placement as the Sgr C NTF. Examining the MeerKAT image in detail, we find that the radio luminosity at the intersection of the east-west ridge of FIR-4 with a few of the two westernmost sub-filaments of the NTF is $\sim$20\% higher than the sum of individual luminosities of the neighbouring regions of the superposed features. Moreover, the potential interaction site displays an increased north-south radio emission width (which is much larger than the MeerKAT beam size; seen in filament \#13 in \citealt{Heywood2022}), and the NTF sub-filaments brighten as they approach the intersection site from both the north and south. These characteristics lead us to suggest that FIR-4 is interacting directly with the Sgr C NTF. 

We note that there is an absence of magnetic field detections in the far-IR dark region that lies north of the Sgr C \hii{} region at latitude -0.02\degree{} and south of the cluster of YSOs. This region has faint emission at both 8 and 24 \micron{}, but gradually becomes negligible towards far-IR and radio frequencies (c.f., Figure~\ref{fig:NTF}). This indicates that this region contains mainly diffuse warm gas, possibly heated by the nearby YSOs. 

\subsubsection{The G359.33-0.15 Filaments}
\label{sec:faintfilament}
\begin{figure*}
    \centering
    \includegraphics[width=\textwidth]{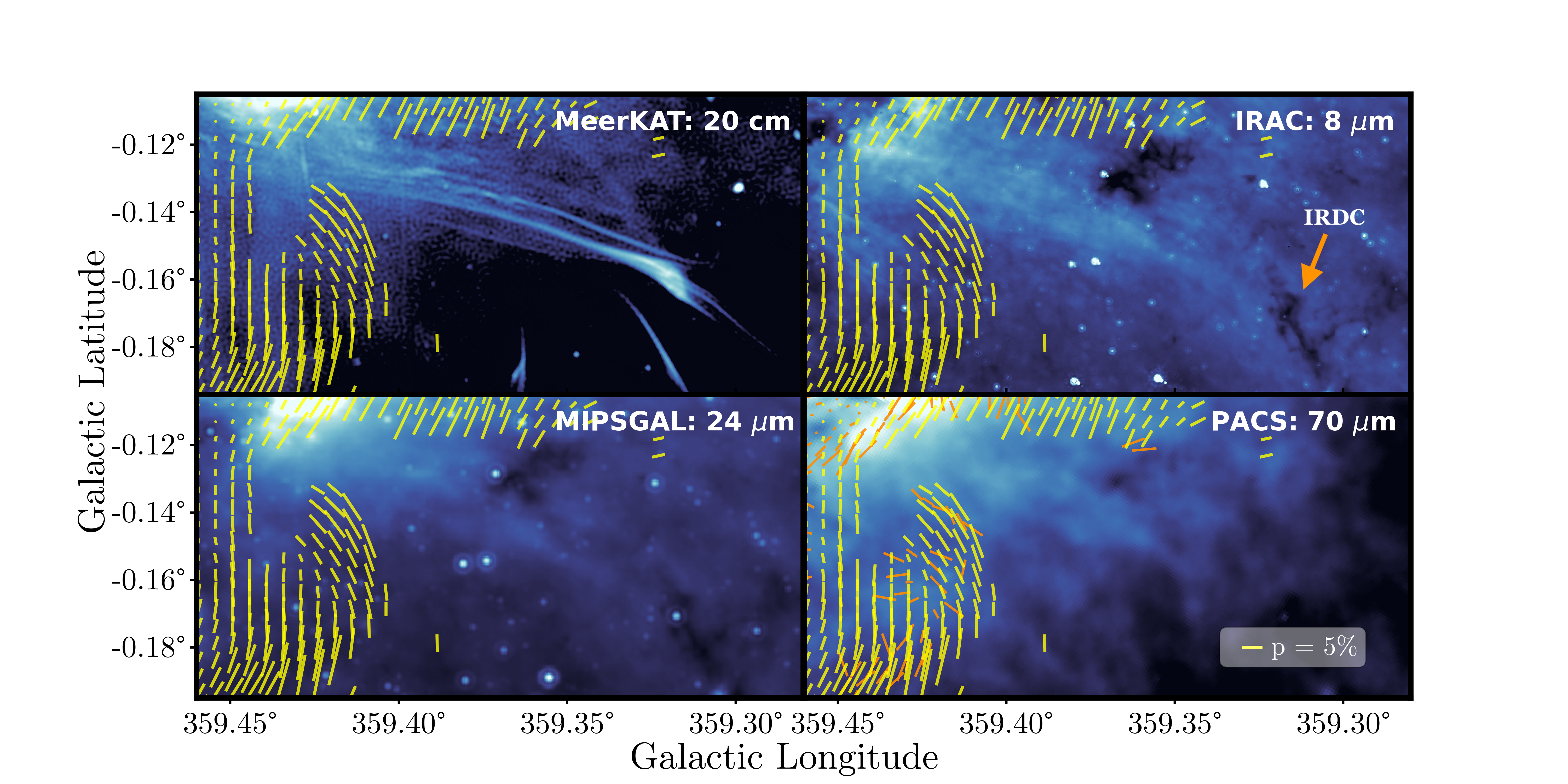}
    \caption{The G359.33-0.15 NTF towards the southwest of the Sgr C \hii{} region shown at 20 cm \citep{Heywood2022, Yusef-Zadeh2022index}, 8 \micron{} \citep{Stolovy2006}, 24 \micron{} \citep{Carey2009}, and 70 \micron{} \citep{Molinari2010}, overlaid with 214 \micron{} magnetic field pseudovectors from \citet{Pare2024}. The bright source in the top left corner all all four panels is the Sgr C \hii{} region. The orange segments in the lower right panel are the 850 \micron{} magnetic field detections from \citet{Lu2024}. The location of an IRDC near $l=359.32$\degree{}, $b=-0.17$\degree{} is labelled by the orange arrow in the top right panel.}
    \label{fig:faintNTF}
\end{figure*}

A fainter bundle of parallel radio filaments, referred to as the ``Hummingbird" by \citet{Yusef-Zadeh2022index}, is located to the southwest of the Sgr C \hii{} region. The northeastern end is coincident with an extended X-ray source noted at energies $<6$ keV (which we detail in Section~\ref{sec:X-ray}). Figure~\ref{fig:faintNTF} shows the MeerKAT/L-Band image of the G359.33-0.15 filament bundle and its possible 8--70 \micron{} counterparts.
Such counterparts would be more characteristic of a thermal source rather than an NTF. However, the MeerKAT spectral index map \citep{Yusef-Zadeh2022index, Bally2024} indicates that the spectrum of at least a portion of the G359.33-0.15 filaments is too steep to be thermal, although the mean spectral index is somewhat ambiguous in this regard. Furthermore, the multiplicity of the sub-filaments is a characteristic commonly found among NTFs. Therefore, the apparent infrared counterparts to the G359.33-0.15 filaments could be coincidental and unrelated. No far-IR emission beyond 70 $\mu$m has been detected toward G359.33-0.15, and consequently, there are no 214 \micron{} magnetic field detections to report along the G359.33-0.15 filaments.  

Unlike most NTFs in the CMZ, which are perpendicular to the Galactic plane, the G359.33-0.15 filaments are much closer to being parallel to the Galactic plane. Another of its notable characteristics is that it is apparently interacting with an infrared dark cloud (IRDC) at its southwestern end (in the top two panels in Figure~\ref{fig:faintNTF} at near $l=359.32$\degree{}, $b=-0.17$\degree{}, and labelled by orange arrow in the top right panel), and that interaction has led to a strong morphological bunching and brightening of the radio emission at the interface (see Fig. 3k of \citet{YZ+22b} for additional radio detail). The interaction has also led to the production of a slightly extended X-ray source at or very near that location \citep[][their designation: G359.32-0.16]{YZ+07a}, raising the possibility that this location is where the relativistic electrons that illuminate the G359.33-0.15 filament are produced, by analogy with G359.1-0.2 \citep{UchidaK+96} or G0.18-0.04, and the Sickle \citep{Serabyn1994}. This interesting structure clearly warrants further investigation. Unfortunately, no magnetic field detections have been measured along the G359.33-0.15 NTF and the IRDC with which it is interacting due to its low far-IR luminosity \citep{Pare2024}.

\subsubsection{Source C: G359.39-0.08}
\begin{figure*}
    \centering
    \includegraphics[width=\textwidth]{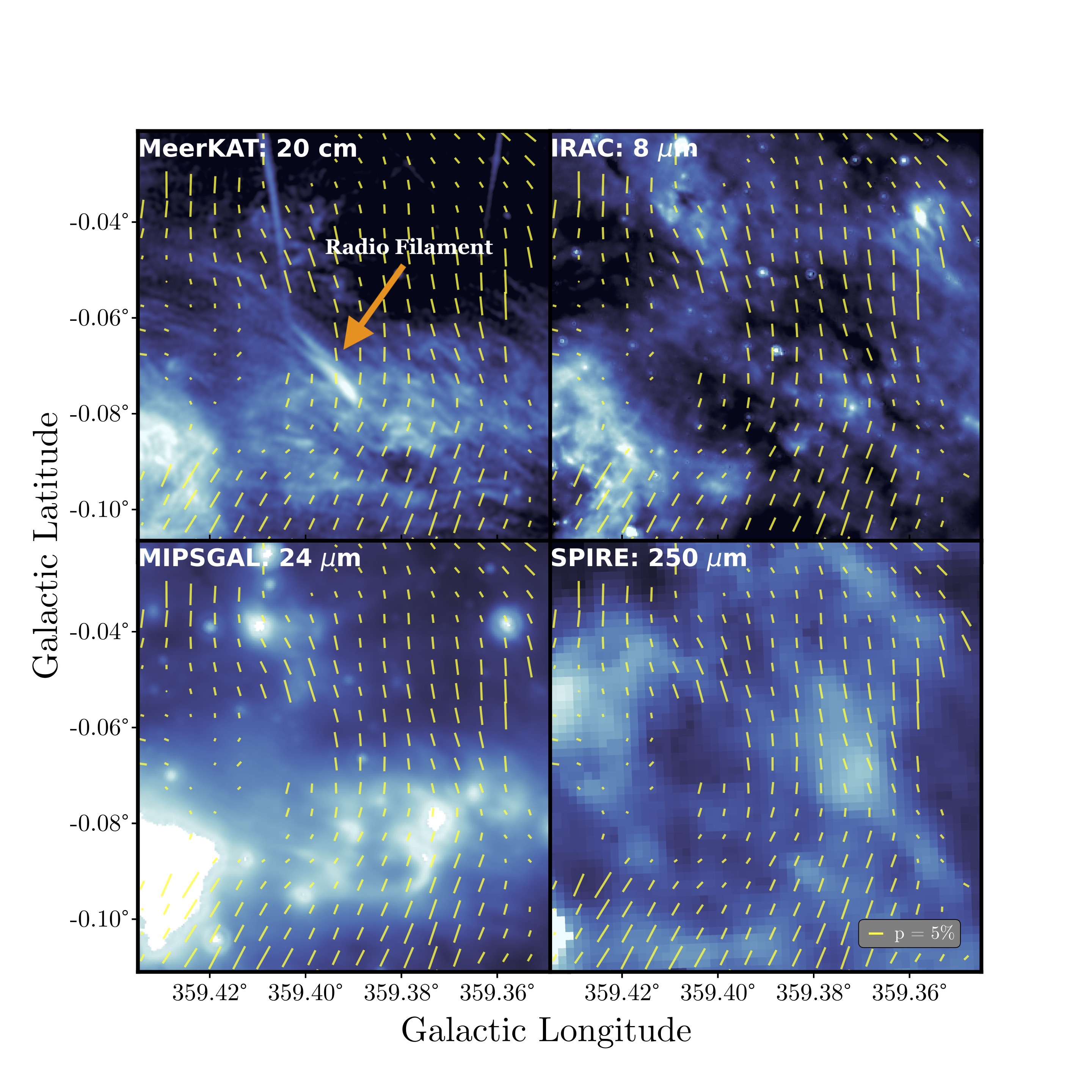}
    \caption{The FIREPLACE 214 \micron{} magnetic field pseudovectors around Source C, superimposed on emission maps at 20 cm \citep{Heywood2022}, 8 \micron{} \citep{Stolovy2006}, 24 \micron{} \citep{Carey2009}, and 250 \micron{} \citep{Molinari2010}. The yellow segments are the 214 \micron{} magnetic field pseudovectors reported by \citet{Pare2024}. No 850 \micron{} magnetic field detections from \citet{Lu2024} were reported near Source C. The orange arrow labels the location of the radio filament near source C. }
    \label{fig:sourceC}
\end{figure*}

Source C is a radio source associated with a diffuse \hii{} region to the immediate west of the Sgr C \hii{} region, where thermal counterparts have been reported \citep{Liszt1995, Hankins2020}. Figure~\ref{fig:sourceC} displays multi-wavelength structures of Source C overlaid with the 214 \micron{} magnetic field pseudovectors, which exhibit a uniformly vertical structure with a similar polarization fraction. 
The magnetic field detections imply a low turbulent-to-ordered ratio and a largely uniform $|{\bf B}_{\rm POS}|$, as shown in Figure~\ref{fig:B_POS}. 

Source C contains a short radio filament with a bright southern end that ends abruptly at $l=359.390$\degree{}, $b=-0.077$\degree{} in the MeerKAT/L-Band image (labelled in the top left panel of Figure~\ref{fig:sourceC}). To the northeast of that point, there are two adjacent and parallel filaments, the upper one of which appears to undergo a sudden northward deflection at $l=359.405$\degree{}, $b=-0.06$\degree{}, above which its brightness increases. The deflection point of this filament is located within a far-IR emission cavity centred at $l=359.405$\degree{}, $b=-0.060$\degree{}, where we find no significant magnetic field detection, but north of the cavity, the magnetic field orientation is approximately aligned with the filament. If the filament is a magnetic NTF, its ``deflection" of the top filament could occur where it transitions going northward from the far-IR cavity into the local molecular cloud, thereby reflecting the differing field orientations in the two media. 

Another notable feature in Source C is the curvature of the magnetic field around the peak of the 24 \micron{} emission at $l=359.37$\degree{}, $b=-0.08$\degree{}. This could perhaps be attributed to the expansion of the ionized gas in this region.

\subsubsection{The Faint Southern Tail of the Sgr C NTF \& Source G359.467-0.17}
\label{sec:radioriver}
\begin{figure*}
    \centering
    \includegraphics[width=\textwidth]{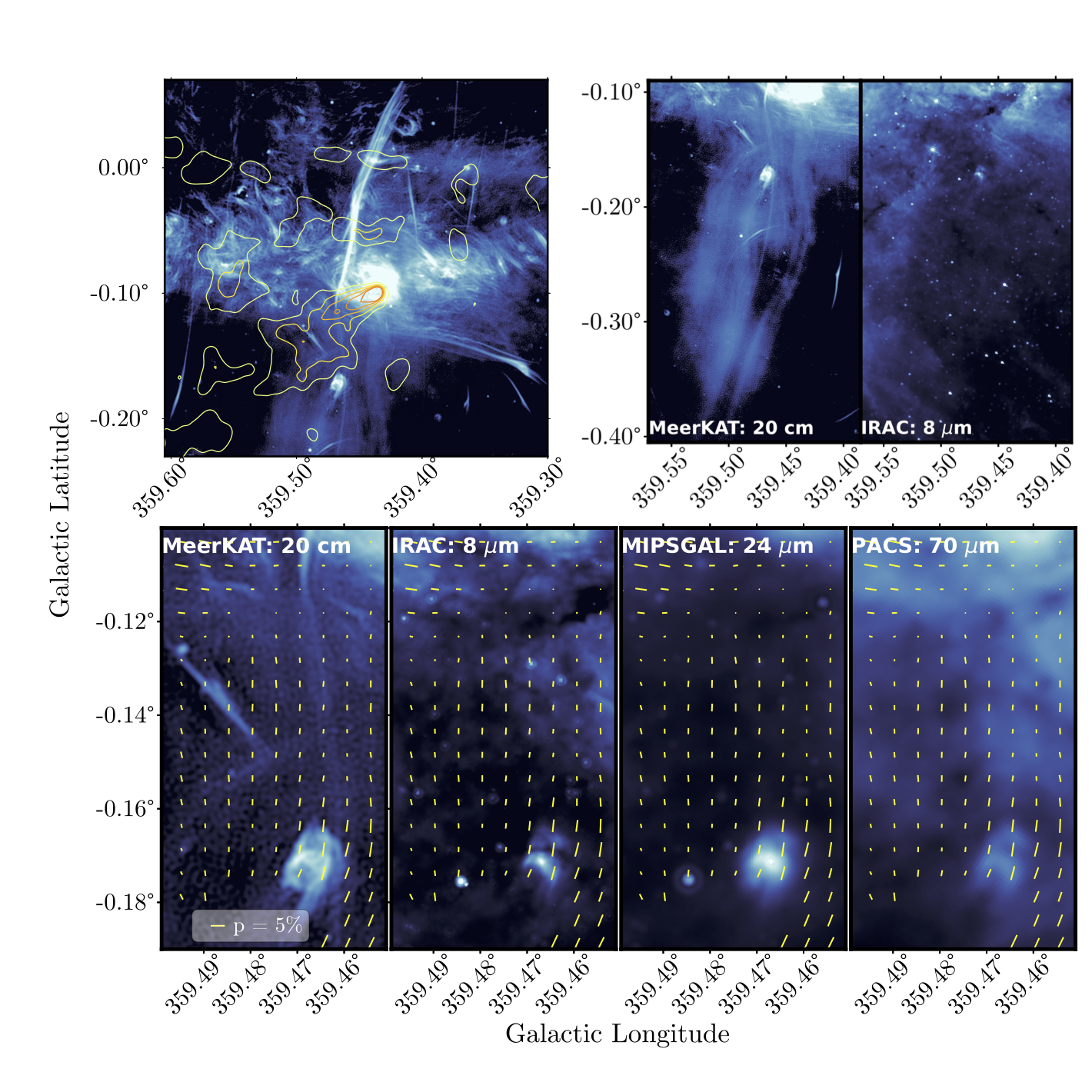}
    \caption{\textit{Top left:} The full extent of the Sgr C NTF in radio continuum \citep{Heywood2022}. The orange contours are the FIREPLACE 214 \micron{} emission at (3, 5.25, 7.5, 9.75, 12) Jy/pixel from \citet{Pare2024}. Note the faint southern extension of the bright NTF bundle to the north. \textit{Top right:} The faint southern ``tail" of the Sgr C NTF, and source G359.467-0.17, displayed at wavelengths of 20 cm \citep{Heywood2022} and 8 \micron{} \citep{Stolovy2006}. \textit{Bottom:} The extended 20 cm radio emission towards the south of the Sgr C \hii{} region, from \citet{Heywood2022}, 8 \micron{} \citep{Stolovy2006}, 24 \micron{} \citep{Carey2009}, and 70 \micron{} \citep{Molinari2010}, overlaid with the 214 \micron{} magnetic field pseudovectors from \citet{Pare2024}.}
    \label{fig:radio}
\end{figure*}

The prominent radio NTF extending north from Sgr C has an apparent faint southern ``tail" that extends onto the $-65$ \kms{} cloud, in projection, starting at $l=359.468$\degree{}, $b=-0.12$\degree{}. The top left panel of Figure~\ref{fig:radio} shows the full extent of the Sgr C NTF, including the faint southern extension.
It is apparent that there is a sharp decline in the brightness of the Sgr C NTF precisely where it overlaps with the head-tail cloud (depicted in yellow contours in the top left panel of Figure~\ref{fig:radio}) on the sky-plane, suggesting that the NTF is interacting with the head-tail cloud. A zoomed-in view of the southern extension is shown in the top right panel of Figure~\ref{fig:radio}. This tail is parallel to a bundle of other radio filaments having a more diffuse nature, shown in the top right panel of Figure~\ref{fig:radio} \citep[see also Figure 12 of][]{Yusef-Zadeh2022index}. Additionally, we note the obvious spatial anti-correlation between this bundle of filaments with the PAH emission shown in 8 \micron{}.

The NTF tail persists until $l=359.47$\degree{}, $b=-0.25$\degree{}. Along its trajectory, the tail is tangent in projection to a compact radio source, G359.467-0.17 (shown in detail in the bottom panel of Figure~\ref{fig:radio}), which has the morphology of a tilted bipolar nebula, along with clear IR counterparts that indicate a centrally heated, dusty nebula. Because the brightness of the NTF tail does not vary as a function of its proximity to G359.467-0.17, it seems unlikely that the two structures are interacting.

\subsection{The Sgr C \hii{} Region}
The 19.6\arcsec{} resolution of the FIREPLACE survey \citep{Pare2024} allows us to closely examine the magnetic field morphology near the Sgr C \hii{} region. From Figure~\ref{fig:214Bfield}, we observe a cavity within the Sgr C \hii{} region, in which resides relatively high-temperature dust, as manifested in mid-IR maps (Panels \#2-4 in Figure~\ref{fig:SgrC9}). In this paper, we focus on magnetic field detections in cold molecular clouds surrounding the \hii{} region. Because such warm regions are not well sampled by our 214 \micron{} survey, the field there would best be studied with future mid-IR polarimetry; indeed, a recent study by \citet{Bally2024} also inferred the existence of strong magnetic fields in this \hii{} region, in the wavelength range 1-5 \micron{} with \textit{JWST}.

The magnetic field orientation in the northern hemisphere of the Sgr C \hii{} region is mostly tangential to the radio shell shown in both Figure~\ref{fig:214Bfield} and Panel \#9 of Figure~\ref{fig:SgrC9}. This tangential morphology along the northern surface of the \hii{} region is consistent with the results of magnetohydrodynamic simulations \citep{Krumholz2007, Arthur2011}. We posit that the expanding shell of the \hii{} region, presumably driven by stellar winds and the advancing ionization front, compresses the tangential component of the field in the surrounding clouds to the point where it has become the dominant component.

However, the magnetic field orientation near the head-tail cloud and the G359.44-0.10 EGO displays a different morphology. Near the EGO, the magnetic field is mainly influenced by activity in the EGO (protostellar winds and jets) instead of the ionization front of the Sgr C \hii{} region, which is behind the head-tail cloud \citep{Lang2010}. We return to a more detailed discussion of the magnetic field morphology of the head-tail cloud in Section ~\ref{sec:headtail}.

\subsubsection{[\cii{}] Emission: A ``Shell" around the \hii{} Region}
\begin{figure}
    \centering
    \includegraphics[width=\columnwidth]{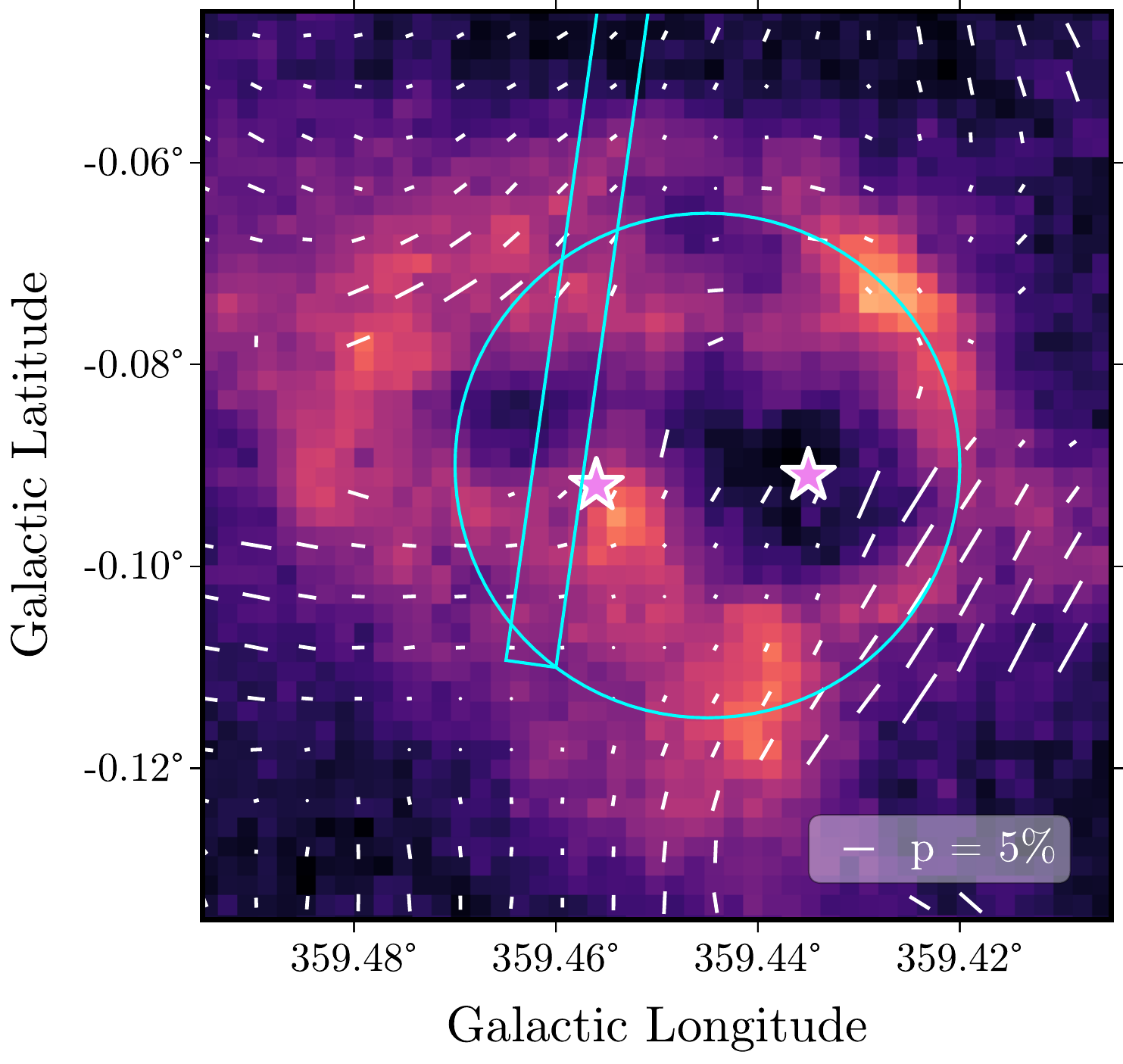}
    \caption{The {\it CMZ in [\cii{}] Survey} image of the $-41$ \kms{} velocity channel observed using SOFIA/upGREAT \citep{Riquelme2025}. White segments are the 214 \micron{} pseudovectors made from data reported by \citet{Pare2024}. Pink stars indicate the positions of two Wolf-Rayet stars reported by \citet{Geballe2019}. The cyan schematics mark the approximate location and extent of the Sgr C \hii{} region and the Sgr C NTF.}
    \label{fig:CII}
\end{figure}

To shed more light on the properties of the Sgr C \hii{} region, we use the [\cii{}] line analysis conducted by \citet{Riquelme2025} using the upgraded German Receiver for Astronomy at Terahertz Frequencies (upGREAT) instrument aboard the SOFIA telescope. The [\cii{}] line traces the warm ionized and partially ionized media exposed to UV photons from the HII region. Figure~\ref{fig:CII} displays the \cii{} velocity channel image at $-41$ \kms{} \citep{Riquelme2025} overlaid by the FIREPLACE 214 \micron{} magnetic field pseudovectors for comparison with the morphology of the \cii{} shell.
This particular velocity channel is chosen to best highlight the cavity structure revealed by the [\cii{}] line. In the immediate vicinity of the Sgr C \hii{} region, the 850 \micron{} magnetic field measurements reported by \citet{Lu2024} are largely consistent with the 214 \micron{} ones shown here. 

Figure~\ref{fig:CII} reveals what appears to be a double-cavity morphology associated with the Sgr C \hii{} region, with the western cavity being larger in size. The [\cii{}] shell surrounds and overlaps with the outer perimeter of the \hii{} region appearing in radio emission (see Panel \#9 in Figure~\ref{fig:SgrC9}), presumably identifiable as the ionization front. The 214 \micron{} magnetic field orientation is mostly tangential to the cavities, except in the southeast direction because of the dense foreground head-tail cloud. The [\cii{}] emission in this shell is found to deviate somewhat from circular symmetry, with the strongest emission in the northwest and near the G359.44-0.10 EGO. Together with the tangential magnetic field morphology, these observations resonate with the conclusion found in the simulations of \citet{Arthur2011}, that the expansion of an \hii{} region creates a tangential magnetic field orientation around its border. Moreover, the [\cii{}] cavities are separated by a ``bridge" of faint [\cii{}] emission centred at $l=359.455$\degree{}, $b=-0.090$\degree{}. Only two significant 214 \micron{} pseudovectors are present toward the ``bridge," and both are parallel to it.

The polarization fraction in high [\cii{}] intensity regions appears low due to their high column density and high turbulent-to-ordered ratio, as shown in Section~\ref{sec:ang_disp}. Indeed, for the same region, the inferred $|{\bf B}_{\rm POS}|$ near the cavity is low. 

\citet{Geballe2019} reported two dusty Wolf-Rayet stars that are located in the Sgr C \hii{} region (pink-filled stars in Figure~\ref{fig:CII}). The western Wolf-Rayet star is positioned at the centre of the western cavity, and is therefore a strong candidate for contributing to the formation of this cavity and to the ionization of the Sgr C \hii{} region. Its winds are potentially responsible for the tangential magnetic field morphology around the rim of the Sgr C \hii{} region. 

\subsubsection{X-rays in the Sgr C \hii{} region} 
\label{sec:X-ray}

\begin{figure}
    \centering
    \includegraphics[width=\columnwidth]{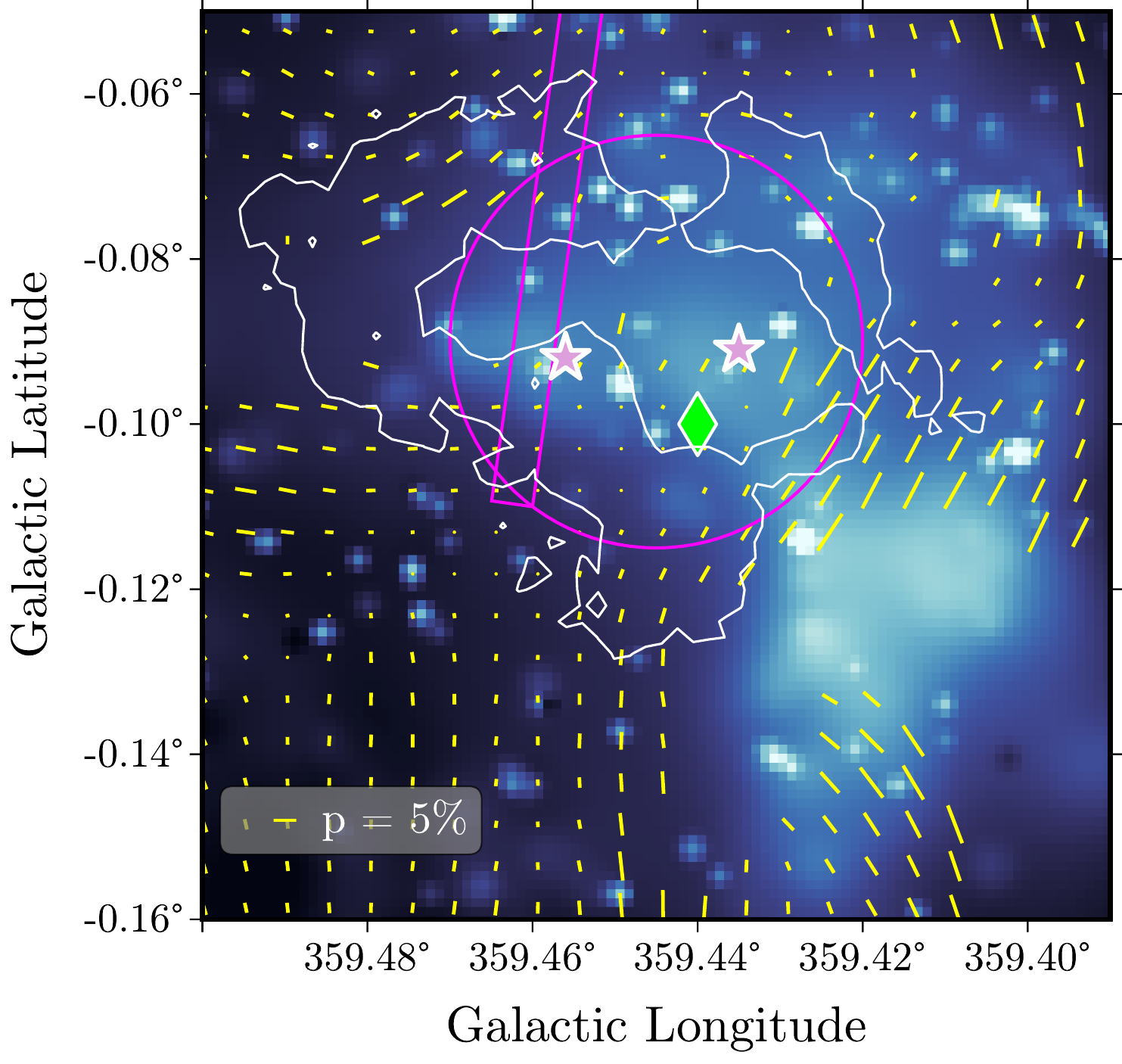}
    \caption{The 2.5--4 keV \textit{Chandra} X-ray image reported by \citet{Wang2021}. The white contour outlines the [\cii{}] emission in the $-41$ \kms{} velocity channel at the intensity of 13 K \citep{Riquelme2025}. The yellow segments are the 214 \micron{} magnetic field pseudovectors from \citet{Pare2024}. The pink-filled star symbols mark the locations of two Wolf-Rayet stars \citep{Geballe2019}. The purple shapes denote the location of the Sgr C \hii{} region and the Sgr C NTF, and the green diamond indicates the location of the EGO at the ``head" of the head-tail cloud. }
    \label{fig:X-ray}
\end{figure}

In projection, Sgr C is associated with an X-ray source having the spectroscopic and luminosity characteristics of a supernova remnant \citep{Tsuru+09}.  However, because the relatively low-resolution Suzaku observations reported by \citet{Tsuru+09} show that the peak X-ray emission in the 2.45 keV S XV K$\alpha$ line is displaced from the Sgr C \hii{} region, those authors conclude that the X-rays are unlikely to emanate from the \hii{} region. A more recent observation by \citet{Wang2021} with \textit{Chandra} resolves a secondary peak centred on the \hii{} region, as shown in Figure~\ref{fig:X-ray}, which overlays the 214 \micron{} magnetic field pseudovectors on the 2.5--4 keV \textit{Chandra} image.
This figure shows that the brightest portions of the secondary X-ray peak fit within the [\cii{}] shell surrounding the \hii{} region (white contour, from \citealt{Riquelme2025}). Fainter X-ray emission extends away from the \hii{} region both toward the northwest and to the southwest, where it extends continuously at a low level to the brightest X-ray peak lying $\sim2$' (5 pc) from the \hii{} region, at $l=359.42$\degree{}, $b=-0.12$\degree{}.  The two peaks of X-ray emission are possibly connected through a ``bridge" at $l=359.425$\degree{}, $b=-0.103$\degree{}, although they could also be slightly overlapping, independent sources at different distances.

If the X-rays from the secondary source arise from within the \hii{} region, they could indicate that a supernova has taken place among the presumed cluster of massive stars that powers the \hii{} region. The relative paucity of X-ray emission towards the southeast in the bands between 2.5 -- 4  keV can perhaps be attributed to absorption by the head-tail cloud situated in front of the southeast portion of the \hii{} region (\citealt{Tsuru+09} also raised this same point to account for the shape of their X-ray contours at this location).

The stronger peak of extended X-ray emission to the southwest of the Sgr C \hii{} region exhibits a strong spatial anti-correlation with far-IR emission and with the presence of significant 214 \micron{} magnetic field detections (see Figure~\ref{fig:214Bfield}).

We note again that the location of the X-ray emission to the southwest of the Sgr C \hii{} region coincides in projection with the northeast ends of the G359.33-0.15 radio filaments (displayed in Figure~\ref{fig:faintNTF} and detailed in Section~\ref{sec:faintfilament}), although it is not yet clear whether there is any physical association between these two features.

\subsection{The Head-tail Cloud (G359.44-0.10) and the $-65$ \kms{} cloud}
\label{sec:headtail}
\begin{figure*}
    \centering
    \includegraphics[width=\textwidth]{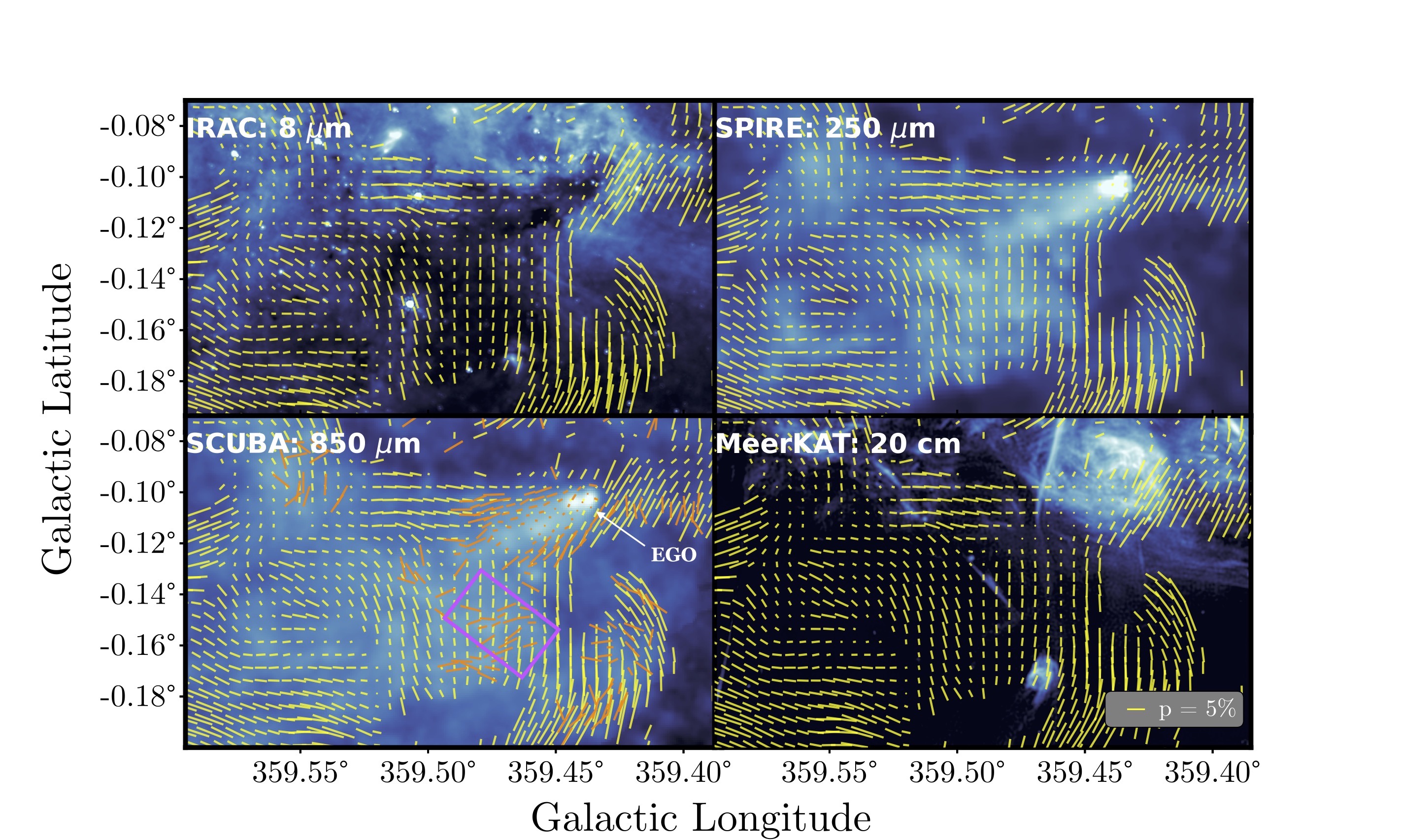}
    \caption{The magnetic field pseudovector representation from FIREPLACE coinciding with the head-tail cloud and the $-65$ \kms{} cloud, displayed at wavelengths of 8 \micron{} \citep{Stolovy2006}, 250 \micron{} \citep{Molinari2010}, 850 \micron{} \citep{Pierce-Price2000}, and 20 cm \citep{Heywood2022}. The orange pseudovectors in the lower left panel are the 850 \micron{} magnetic field pseudovectors from \citet{Lu2024}. The purple rectangle in the lower left panel denotes the extent of source AGAL 359.474-0.152 detected in the 870 \micron{} ATLASGAL survey \citep{Contreras2013}. The white arrow in the lower left panel labels the location of the EGO. }
    \label{fig:headtail}
\end{figure*}

To the southeast of the Sgr C \hii{} region lies the head-tail cloud. The head-tail cloud consists of two parts: an exceptionally bright compact source (the G359.43-0.10 EGO or the ``head"; \citealt{Kendrew2013, Crowe2024}; green star in the lower left panel of Figure~\ref{fig:headtail}) and a diffuse and cometary ridge of high far-IR luminosity, high column density (see Figure~\ref{fig:colden}), and low polarization fraction extending towards the southeast (the ``tail" addressed in Section~\ref{sec:radioriver}). This cloud is also found to exhibit a consistently high turbulent-to-ordered ratio ($\langle {\bf B}_t^2\rangle/\langle {\bf B}_0^2\rangle$), as detailed in Section~\ref{sec:ang_disp}. Figure~\ref{fig:headtail} shows the magnetic field pseudovectors of the head-tail cloud and the $-65$ \kms{} cloud overlaid with various background images. 
Initially discovered by \citet{Forster2000} and \citet{Yusef-Zadeh2009}, \citet{Kendrew2013} confirm the existence of early-stage star formation within the EGO, driving multiple jet-like protostellar outflows, recently detailed by \citet{Crowe2024}. These magnetic field orientations can be due to the cloud being sheared along its long dimension. Alternatively, the head-tail cloud is moving as a unit with constant velocity, and the magnetic field lines threading through that cloud, originally roughly perpendicular to the cloud, can be deformed and dragged into their observed orientations by the bulk motion of the cloud, into which the field lines are frozen. Towards the southeast of the EGO, the tail forms a coherent structure continuous with the $-65$ \kms{} cloud, as depicted in Figure~\ref{fig:HCN_vel}. 

The JCMT/POL2 polarimetric study surveyed this region at 850 \micron{} (\citealt{Lu2024}; shown in orange in the lower left panel of Figure~\ref{fig:headtail}). We note that the 850 \micron{} magnetic field pseudovector orientations near the ``head" generally agree with those of \citet{Pare2024}. However, in the body of the $-65$ \kms{} cloud near $l=359.47$\degree{}, $b=-0.16$\degree{} (purple rectangle in the lower left of Figure~\ref{fig:headtail}), there are several 850 \micron{} magnetic field detections which appear roughly parallel to the Galactic plane and perpendicular to those detected at 214 \micron{}. This misalignment could occur if the polarimetric surveys at different wavelengths are detecting emission from different temperature regimes within the clouds or different components of the same cloud along a given line of sight. 

\begin{figure*}
    \centering
    \includegraphics[width=\textwidth]{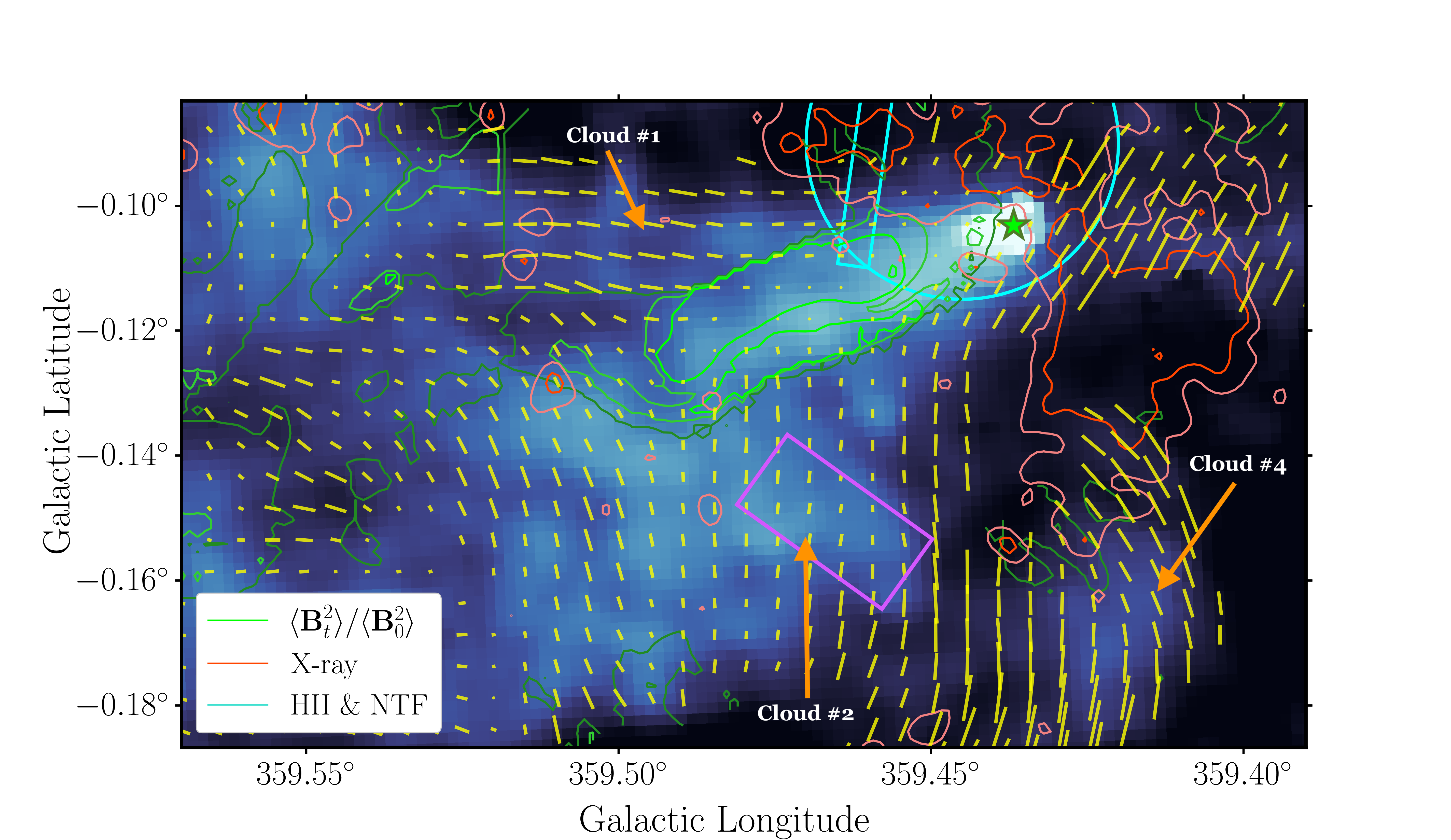}
    \caption{Various morphological features within the $-65$ \kms{} cloud. The yellow segments are the 214 \micron{} magnetic field pseudovectors from \citet{Pare2024}, superposed upon the {\it Herschel}/SPIRE 250 \micron{} intensity image. The green contours are the $\langle {\bf B}_t^2\rangle/\langle {\bf B}_0^2\rangle$ ratio at values of (0.1, 0.6, 1.2). The red contours depict the 2.5-4 keV X-ray intensity from \citet{Wang2021} at (5000, 7000) ${\rm counts\,s^{-1}\,arcmin^{-2}}$. The cyan outlines indicate the extent of the Sgr C \hii{} region and the Sgr C NTF. The purple rectangle denotes the extent of source AGAL 359.474-0.152 detected in the 870 \micron{} ATLASGAL survey \citep{Contreras2013}. The green star labels the location of the EGO. The orange arrows label the extent of clouds \#1, \#2, and \#4 in Figure~\ref{fig:masks} used in the DCF analysis in Section~\ref{sec:DCF}. }
    \label{fig:headtail_morpho}
\end{figure*}

Furthermore, we outline a few intriguing morphological associations between the 214 \micron{} magnetic field pseudovectors and several observed structures in the $-65$ \kms{} cloud, shown in Figure~\ref{fig:headtail_morpho}.
The magnetic field within the $-65$ \kms{} cloud is separable into three distinct parts, whose spatial extent is well represented by regions \#1, \#2, and \#4 of the $-65$ \kms{} cloud in Figure~\ref{fig:masks}. In this subsection, we continue to adopt this naming scheme for our discussion, and in the following, we summarize the magnetic field morphology of each region, and describe potential relationships with structures observed at other wavelengths: 
\begin{itemize}
    \item \textbf{Cloud \#1.} This cloud, centred at $l=359.495$\degree{}, $b=-0.11$\degree{}, lies between the Galactic plane to the north (the far-IR-dark region near the top of Figure~\ref{fig:headtail}) and the high turbulent-to-ordered ratio ``ridge" (the green contours in Figure~\ref{fig:headtail}) in the south. \citet{Lu2024} argue that the field morphology in this region, which is parallel to the Galactic plane, is due to its interaction with numerous \hii{} regions along the Galactic plane \citep{Carey2009, Hankins2020}. 
    \item \textbf{Cloud \#2.} This cloud, centred at $l=359.47$\degree{}, $b=-0.15$\degree{}, is located to the south of Cloud \#1, bordered by the high turbulent-to-ordered ratio ``ridge" to the north and the 2.5-4 keV X-ray emission source to the west (orange contours in Figure~\ref{fig:headtail_morpho}). The magnetic field within Cloud \#2 is uniformly perpendicular to the Galactic plane and to the magnetic field orientation observed in Cloud \#1. However, the magnetic field transition from Cloud \#1 to Cloud \#2 takes place toward the high turbulent-to-ordered ratio ``ridge," thus obscuring the transition from our perspective. We note additionally that the SiO emission in Figure~\ref{fig:SiO_vel} is exclusively found in Cloud \#2, likely supporting the case that the magnetic field systems in these listed clouds are segregated.
    \item \textbf{Cloud \#4.} Centered at $l=359.425$\degree{}, $b=-0.17$\degree{}, Cloud \#4 is separated from Cloud \#2 above $b=-0.15$\degree{} but merges with Cloud \#2 below this latitude. We note that the far-IR ``gulf" approximately located at $l=359.43$\degree{}, $b=-0.14$\degree{} exhibits an intriguing spatial anti-correlation with the 2.5-4 keV X-ray emission (orange contours in Figure~\ref{fig:headtail_morpho}). \citet{Pare2024} argues that the curved magnetic field morphology of Cloud \#4 can possibly be ascribed to winds associated with winds associated with star formation in the aforementioned 870 \micron{} source AGAL 359.474-0.152 \citep{Contreras2013}. 
\end{itemize}
We note that the observation we provided in this section only pertains to the 214 \micron{} magnetic field measurements. For instance, we do not take into account the observed JCMT/POL2 pseudovectors in Cloud \#2 \citep{Lu2024}. However, we note again that the JCMT/POL2 measurements may be sampling from a different temperature regime or another cloud along the same line of sight. A multi-wavelength dust polarimetry measurement that covers a larger area is warranted to properly interpret the observational differences at the two wavelengths.

\section{Conclusion}
\label{sec:conclusion}
This paper presents a detailed analysis of the FIREPLACE magnetic field measurements from the FIREPLACE dust polarimetry survey \citep{Pare2024} in the Sgr C complex.

We summarize our findings as follows:
\begin{itemize}
    \item As was reported in the earlier FIREPLACE papers \citep{Butterfield2024, Pare2024, Pare2024b}, We find that the polarization fraction exhibits an anti-correlation with 214 \micron{} intensity. As such, the polarization fraction on the boundary of clouds is typically higher than in the centre of clouds. The higher polarization fraction is likely a result of a combination of two effects: 1) decreased turbulence towards the boundary of clouds, as indicated by our angular dispersion analysis and 2) by the likelihood that the multiplicity of different magnetic domains within the cloud is likely to be smaller toward the edges of clouds.  
    \item Through a modified DCF analysis of individual clouds, we find the sky-plane magnetic field strength to be $30~\mu{\rm G} - 1~{\rm mG}$ in Sgr C. Regions with higher turbulence typically have weaker magnetic field strength, such as the head-tail cloud and near the Sgr C \hii{} region. The dominant determinant of the plane-of-sky magnetic field strength is the local turbulence. 
    \item The expansion of the Sgr C \hii{} region causes extensive turbulent behaviour in its vicinity, as shown by our DCF analysis. We find that the \hii{} region's interior has been cleared of cold dust content since no far-IR emission is seen there, thereby creating a cavity devoid of significant 214 \micron{} magnetic field pseudovectors. Additionally, we find that the 214 magnetic field orientation is largely tangential to the [\cii{}] shell surrounding the \hii{} region reported by \citet{Riquelme2025}, and briefly shown in Figure~\ref{fig:CII}. The angular dispersion analysis again confirms the elevated turbulence in this shell, accompanied by existing molecular shock tracers. We note an intriguing diffuse source of X-ray emission to the southwest of the Sgr C \hii{} region, which warrants further study as potentially resulting from past activity in the Sgr C region.
    \item The magnetic field morphology near the base of the Sgr C NTF is consistent with the hypotheses that either magnetic field line reconnection or diffusive shock acceleration at the ionization/compression front of the \hii{} region is responsible for the relativistic electrons that illuminate the NTF with their synchrotron emission. Also, the field direction in the $-90$ \kms{} cloud is found to be parallel to the filament. 
    \item We have examined the morphological correspondences between the 214 and 850 \micron{} magnetic field pseudovectors and numerous radio features, such as Source C, FIR-4, and the G359.33-0.15 filaments. From these comparisons, we conclude that many radio structures in the Sgr C region exhibit morphological correspondences with the observed magnetic field direction. Future multi-wavelength dust polarimetry surveys, coupled with detailed molecular line surveys, are needed to elucidate the dynamics of these structures and to determine how the magnetic field and the gas are interacting. 
\end{itemize}

\section{Acknowledgements}
This work is primarily based on observations made with the NASA/DLR Stratospheric Observatory for Infrared Astronomy (SOFIA). SOFIA was jointly operated by the Universities Space Research Association, Inc. (USRA), under NASA contract NNA17BF53C, and the Deutsches SOFIA Institut (DSI) under DLR contract 50 OK 2002 to the University of Stuttgart. 

Financial support for this work was provided by NASA through award \#09-0054 through USRA. This work has made extensive use of NASA's Astrophysics Data System (\href{http://ui.adsabs.harvard.edu/}{http://ui.adsabs.harvard.edu/}) and the arXiv e-Print service (\href{http://arxiv.org}{http://arxiv.org}). In addition, part of this work is completed using the computing resources provided by the University of Chicago’s Research Computing Center. 

RJZ and MRM are supported by NASA through award \#09-0054 administered as a subgrant to UCLA. RJZ thanks Keila Cunha e Silva and Joan-Emma Shea for generously sharing their computing resources, and Q. Daniel Wang for kindly providing his data. Finally, RJZ wishes to acknowledge all community members of Meadowridge School, located in British Columbia, Canada, for their home-like hospitality and helpful conversations while much of this work was completed. 

The authors at UCLA Department of Physics \& Astronomy acknowledge our presence on the traditional, ancestral, and unceded territory of the Gabrielino/Tongva peoples. Much of this work was completed at Meadowridge School, located on the ancestral and unceded territories of the Katzie, Kwantlen, and Coast Salish Peoples, as well as the University of Chicago, located on the ancestral and unceded territories of the Kickapoo, Peoria, Potawatomi, Miami, and Sioux people. We value the opportunity to learn, live, and share research and educational experiences on these traditional lands.

\facility{SOFIA, {\it Chandra}, {\it Herschel}, JCMT, MeerKAT, Mopra, {\it Spitzer}}
    
\software{\textsc{Astropy} \citep{Astropy2013, Astropy2018}, \textsc{CASA} \citep{CASA}, \textsc{cmocean} \citep{cmocean}, \textsc{emcee} \citep{Foreman-Mackey2013}, \textsc{Matplotlib} \citep{Matplotlib}, \textsc{Numpy} \citep{Numpy}, \textsc{Pandas} \citep{Pandas}, \textsc{polBpy} \citep{PolBpy}, \textsc{Scipy} \citep{SciPy}}

\appendix
\counterwithin{figure}{section}
\counterwithin{table}{section}

\renewcommand{\thesection}{A.\arabic{section}}
\renewcommand{\thefigure}{A.\arabic{figure}}
\renewcommand{\thetable}{A.\arabic{table}}
\addtocounter{table}{-1}

\section{Autocorrelation Function of Sgr C Clouds}
\label{appx:A}
\begin{figure*}
    \centering
    \includegraphics[width=\textwidth]{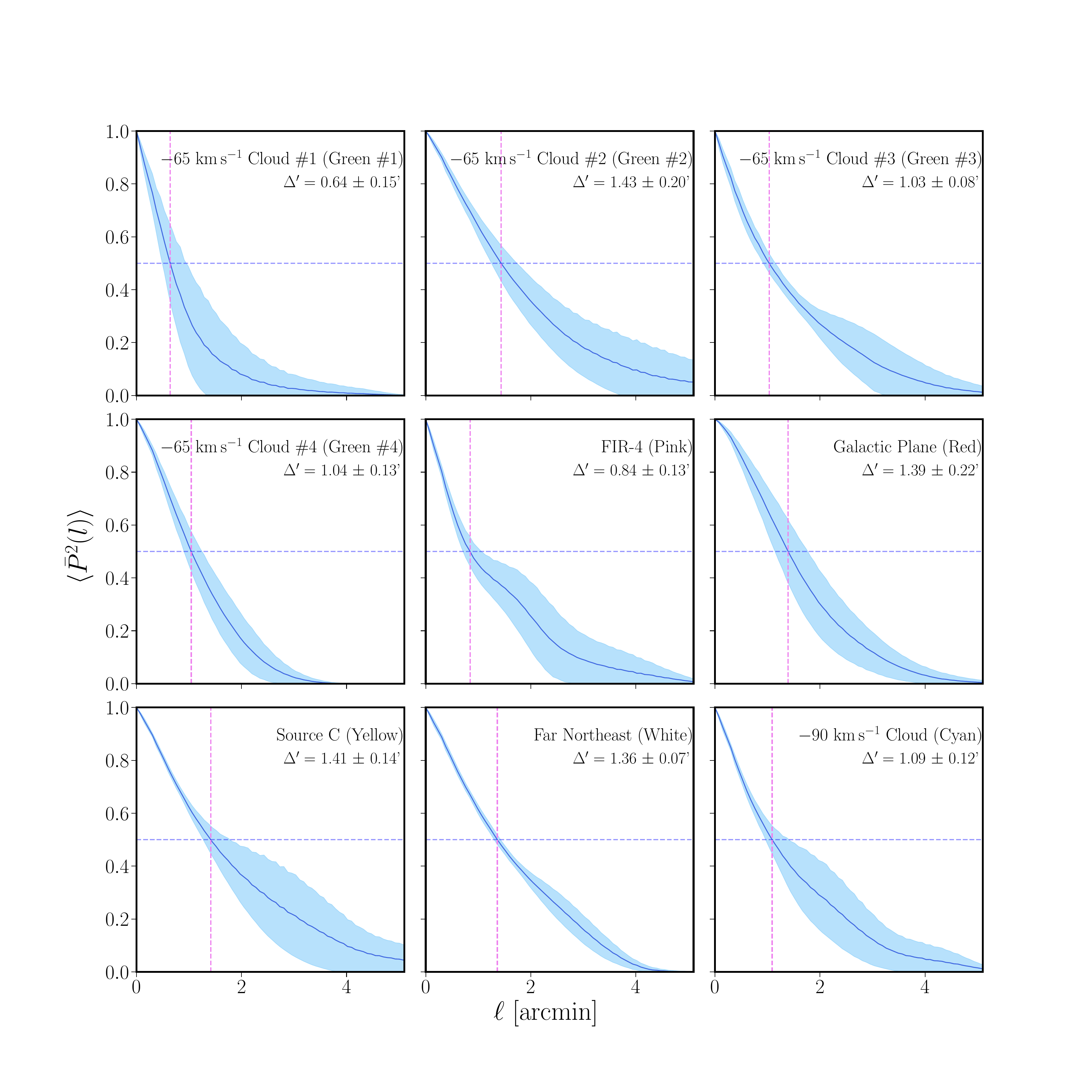}
    \caption{The autocorrelation functions computed from individual Sgr C clouds, as defined in Figure~\ref{fig:masks}. The blue curves are the autocorrelation functions ($\langle\bar{P}^2(l)\rangle$), and the shaded region indicate the standard error. The purple horizontal line indicates where $\langle\bar{P}^2(l)\rangle=0.5$. The burgundy vertical line indicates the HWHM value of the autocorrelation function, which is equivalent to the effective cloud depth. }
    \label{fig:autocor}
\end{figure*}

In this section, we present more details of the autocorrelation function computation and effective cloud depth. We show the shapes of the autocorrelation function calculated for the individual Sgr C clouds in Figure~\ref{fig:autocor}. 
We refer the readers to Figure~\ref{fig:masks} for the location and extent of individual regions. The cloud depth is positively correlated with the size and uniformity of a region. The autocorrelation functions typically exhibit a monotonic decline, except for FIR-4 where a ``bump" near $l\approx1.4$\arcmin{} is observed (central panel). The shape of the FIR-4 autocorrelation function is likely a result of its morphology. The vertical extent of FIR-4 is roughly 1.4\arcmin{} or 0.02\degree{}, therefore the autocorrelation function likely displays the increase of polarized flux towards the northern and southern periphery of FIR-4. 

\bibliography{refs}{}
\bibliographystyle{aasjournal}

\end{document}